\documentclass[preprint,9pt,nopreprintline]{elsarticle} 

\usepackage{etoolbox}

\makeatletter
\newcommand{\ForceMyLayout}{%
  \setlength{\textwidth}{6.0in}
  \setlength{\oddsidemargin}{0.25in}
  \setlength{\evensidemargin}{0.25in}

  \setlength{\textheight}{8.0in}
  \setlength{\topmargin}{0.0in}
  \setlength{\footskip}{40pt}

  \setlength{\hsize}{\textwidth}
  \setlength{\linewidth}{\textwidth}
  \setlength{\columnwidth}{\textwidth}
}
\AtBeginDocument{\ForceMyLayout}
\AtEndEnvironment{frontmatter}{\ForceMyLayout}
\makeatother

\usepackage{amssymb}
\usepackage{amsmath}
\usepackage{hyperref}

\usepackage{booktabs}
\usepackage{chemformula}
\usepackage{wrapfig}
\usepackage{float}
\newcommand{\sbt}{\,\begin{picture}(-1,1)(-1,-3)\circle*{3}\end{picture}\ } 

\usepackage{etoolbox} 
\newtoggle{enablecoms}
\toggletrue{enablecoms} 
\usepackage{xcolor}
\usepackage[normalem]{ulem}
\definecolor{red}{rgb}{1.0, 0, 0}
\definecolor{blue}{rgb}{0, 0, 1.0}
\definecolor{purple}{rgb}{0.54, 0.17, 0.89}
\definecolor{lblue}{rgb}{0, 0.6, 0.83}
\definecolor{green}{rgb}{0, 0.4, 0}

\iftoggle{enablecoms}{
    \newcommand{\com}[1]{{\color{red} [#1]}}

   }{
    \newcommand{\com}[1]{{\color{red} }}

   }

\newcommand{\un}[1]{$\mathrm{#1}$}
\newcommand{\sub}[1]{_{\mathrm{#1}}}

\newcommand{\dd}[1]{\mathrm{d}#1}

\begin{document}


\begin{frontmatter}




\title{A Review of Theory and Practical Considerations of Tunable Diode Laser Absorption Spectroscopy Diagnostics}

\author[label1]{Jose Guerrero}
\author[label1]{Mirko Gamba}
\affiliation[label1]{organization={University of Michigan},
            addressline={1320 Beal Ave},
            city={Ann Arbor},
            postcode={48109},
            state={MI},
            country={United States}}



\begin{abstract}

Tunable Diode Laser Absorption Spectroscopy (TDLAS) has emerged as a versatile and reliable diagnostic tool for measuring temperature, pressure, gas composition, and velocity in power generation and propulsion systems. This paper provides a comprehensive review of TDLAS principles and practical considerations for sensor design and implementation. The discussion begins with a mathematical introduction to the theory of gas absorption including: lineshape modeling and broadening mechanisms, quantitative measurements and challenges, and practical line selection rules. The analysis progresses to wavelength-modulation spectroscopy (WMS), highlighting its advantages in noise rejection and robustness in harsh environments. Furthermore, the calibration-free WMS model and the connection between WMS harmonics and lineshape derivatives is derived. Quantitative measurements through use of multiple harmonics is discussed and challenges surrounding measurement rate are presented. The end of the discussion focuses on practical aspects regarding the implementation of scanned-WMS sensors including laser characterization, background subtraction, and hardware debugging. \\\\
\textbf{Keywords}: tunable diode laser absorption spectroscopy, wavelength-modulation spectroscopy,\\ TDLAS, WMS
\end{abstract}







\end{frontmatter}




\AtEndEnvironment{frontmatter}{\ForceMyLayout}

\section{Introduction}

Tunable Diode Laser Absorption Spectroscopy (TDLAS) has become a standard method for measuring temperature, pressure, and gas composition in many applications across various branches of science and engineering. These include fundamental combustion science \cite{goldenstein2017infrared, bolshov2015tunable}, aerospace propulsion \cite{goldenstein2017infrared}, the automotive industry \cite{rieker2007rapid, stiborek2023mid}, power generation \cite{sun2013tdl}, material science \cite{ikeda1997effects, herman2003optical}, environmental science \cite{srivastava2018development}, atmospheric sciences \cite{werle1993limits, dong2016compact}, plasma physics \cite{ropcke2006application, korolov2021energy}, nuclear engineering \cite{jacquet2013laser, hull2022combined}, the medical industry \cite{wang2009breath, henderson2018laser}, and astronomy and space exploration \cite{mahaffy2012sample, webster2013low, mahaffy2013abundance, minesi2024carbon}. Compared to other laser diagnostic techniques, TDLAS is relatively inexpensive; the cost of a laser diode is only a fraction of the cost of a high-speed video camera required for other diagnostic techniques, such as planar laser-induced fluorescence or schlieren imaging. Additionally, TDLAS requires minimal to no calibration, depending on the specific technique used. It provides high-speed, quantitative, \emph{in-situ} measurements and only requires an optical access point. TDLAS also offers selective species detection, even at very low concentrations. 

Several books \cite{siegman1986lasers, hanson2016spectroscopy, banwell2017fundamentals, eckbreth2022laser} and review papers \cite{goldenstein2017infrared, bolshov2015tunable, liu2019laser} have summarized the theory, key advancements, applications, and state-of-the-art TDLAS diagnostic techniques. For a broader review of laser diagnostic techniques and sensors demonstrated across various applications, readers are directed to the aforementioned review papers. The focus here is to present the theory of absorption by gases in a pedagogical manner, while also addressing practical considerations for the design and implementation of TDLAS sensors. Additionally, this manuscript aims to consolidate many important results and theory published across numerous articles and books into a single cohesive resource. It is intended to serve as a starting reference for first-year graduate students in engineering and researchers new to the field who want to develop an in-depth understanding of TDLAS diagnostics.    

The manuscript is organized into two parts. The first part focuses on the general theory of TDLAS with an emphasis on the Beer-Lambert Law, lineshape functions and broadening mechanisms, quantitative measurements and challenges, uncertainty quantification, and guidelines for line pair selection. The discussion begins by deriving the Beer-Lambert Law using classical optics, which propels the discussion on lineshapes as it establishes a connection to the electric susceptibility $\chi(\omega)$. The Lorentzian lineshape and associated broadening mechanisms are then derived using the classical electron oscillator (CEO) model for atomic/molecular transitions. Although the interaction between electromagnetic radiation and atoms/molecules is fundamentally described by quantum mechanics, it can still be correctly described by the CEO model by expressing the derived equations in terms of measurable quantities \cite{siegman1986lasers}. The purpose of including this discussion is to derive the lineshape models used in laser absorption spectroscopy, as well as their dependence on gas properties, which ultimately enable quantitative measurements. This connects absorption spectroscopy as described in research literature to what can be found in classical optics \cite{born2013principles} or lasers \cite{siegman1986lasers} textbooks. This section is naturally the most math intensive; however, the remaining discussions in part I. of the manuscript are more practical focusing on quantitative measurements for scanned-direct absorption spectroscopy techniques.

The second part examines the TDLAS technique known as wavelength-modulation spectroscopy (WMS). Often the most challenging part for new researchers embarking on implementing WMS sensors is distinguishing among the various WMS techniques available \cite{liu2004wavelength, li2006extension, rieker2009calibration, goldenstein2014fitting, goldenstein2014high, yang2019wavelength, zhu2021second}, and understanding laser characterization methods. Therefore, the introduction to part II. focuses on explaining the advantages and differences of WMS techniques developed in the last 10-15 years. We also dedicate significant attention to deriving the calibration-free WMS model \cite{rieker2009calibration} which is used to simulate WMS signals regardless of which technique is used and thoroughly cover laser characterization strategies for both fixed- and scanned- wavelength-modulation spectroscopy techniques at both kHz-rate and MHz-rate modulation frequencies with more detail than found in other publications. General equations for the $X_{nf}$ and $Y_{nf}$ components of the WMS-${nf}$ harmonic signal derived using a second-order characterization of the laser's intensity are also provided, which previously have only been published for discrete harmonics such as the WMS$-1f$, $-2f$, and $-4f$ signals \cite{rieker2009calibration, mathews2020near}. 

In contrast to more traditional reviews that cover WMS techniques and applications by summarizing the referenced literature, here we attempt to analyze the equations to provide further insights in addition to what is in the referenced literature. This is done for example by providing a discussion on WMS harmonics and their resemblance to lineshape derivatives and by including a discussion to highlight the challenges in achieving high-bandwidth scanned-WMS sensors, which in the last 5 years has been overcome by using bias-tee circuitry. The remaining discussion in part II. details quantitative measurements, guidelines for sensor design, debugging of hardware critical in MHz-rate sensors, and background subtraction. 

To complement the following discussions, a GitHub repository has been made publicly available containing MATLAB and Python code for the simulation of molecular absorption spectra using the HITRAN database \cite{GitHub_Jose}. Also included is code for the simulation of WMS harmonics using the calibration-free WMS model, as well as examples with data for the extraction of WMS harmonics and for laser characterization.

Before proceeding, we provide a brief review of units and nomenclature. In the field of absorption spectroscopy, researchers have adopted the unit of wavenumbers [\un{cm^{-1}}] which is the number of wavelengths ($\lambda$) that fit into a length of one centimeter. Therefore, the unit of wavenumbers describes a spatial frequency $\nu [\mathrm{cm^{-1}}] = 10^7 / \lambda [\mathrm{nm}]$. For conciseness, researchers omit to include the word "spatial" and simply refer to $\nu$ as the frequency or optical frequency of the light. This should not be confused with the temporal optical frequency of the light $f$ which is related to the spatial frequency by $f = c \nu$ where $c$ is the speed of light. In this manuscript, whenever we refer to the frequency of light we will always be referring to the spatial frequency in [\un{cm^{-1}}]. 

The variable $f$ will be reserved for frequency in [\un{Hz}] to avoid confusion when discussing WMS techniques where the frequency of light ($\nu$) is now modulated by a sinusoidal signal of frequency $f$. Another frequency that comes up in this manuscript is $\omega = 2\pi f$, which is an angular frequency in [\un{rad/s}]. This comes up when discussing homogeneous broadening since $\omega$ is standard notation in classical optics textbooks. In any case, all definitions are proportional to each other $\nu \propto f \propto \omega$. Lastly, energy is often reported in units of wavenumbers as well. This simply means that the difference in energy associated with the transition is equivalent to the energy of a photon with frequency $\nu$. The conversion is given by $\Delta E = hc\nu$ where $h$ is Planck's constant. A summary of all frequencies and their units is provided in Table \ref{tab: units}.

\begin{table}[t]
  \centering
  \caption{Summary of variables and units used in NIR laser absorption spectroscopy.}
    \begin{tabular}{p{5em}p{15em}p{5em}}
    \toprule
    \bf Variable & \bf Description & \bf Units \\
    \midrule
    $\lambda$ & wavelength & $\mu$m, nm \\
    $\nu = 1/\lambda$    & wavenumber, optical frequency & \un{cm^{-1}} \\
     $f = \omega/2\pi$  &  frequency & kHz,  MHz \\
    \bottomrule
    \end{tabular}%
  \label{tab: units}%
\end{table}%

{\begin{center}
    \bf Part I -- Theory of Absorption by Gases
\end{center}}

\section{Introduction to Absorption Spectroscopy}

In laser absorption spectroscopy (LAS) experiments, a laser beam with wavelength targeting a specific molecular transition is directed across an absorbing volume of gas and the transmitted light intensity ($I\sub{t}$) is measured by a photodetector. The absorption of light with optical frequency $\nu$ is related to the incident ($I_0$) and transmitted light intensities through Beer-Lambert's Law given by Eq.~(\ref{eq: Beer}).
\begin{equation}\label{eq: Beer}
    \frac{I\sub{t}(\nu)}{I_0(\nu)} = \mathrm{e}^{-\alpha(\nu)}
\end{equation}
The spectral absorbance $\alpha(\nu)$ is further related to the thermodynamic state of the gas through Eq.~(\ref{eq: alpha}).
\begin{align}\label{eq: alpha}
    \alpha(\nu) &= \sum_j \int_0^L S_j(T)\phi_j(\nu-\nu\sub{o},\Delta\nu\sub{c},\Delta\nu\sub{d})PX_i \mathrm{d}\ell \\
    &= \sum_j \int_0^L S_j^*(T)\phi_j(\nu-\nu\sub{o},\Delta\nu\sub{c},\Delta\nu\sub{d})n_i \mathrm{d}\ell \nonumber
\end{align}
The two forms of Eq. (\ref{eq: alpha}) are based on the two conventions for expressing the linestrength function.
The top equation in Eq. (\ref{eq: alpha}) expresses the spectral absorbance in terms of a pressure normalized linestrength $S_j(T)$ \un{[cm^{-2}/atm]}. In this case absorbance is proportional to partial pressure $P_i = PX_i$ \un{[atm]}, where $P$ \un{[atm]} is the gas pressure, and $X_i$ is the absorbing species mole fraction. $T$ \un{[K]} is the gas temperature and $L$ \un{[cm]} is the path length. The bottom equation in Eq. (\ref{eq: alpha}) expresses the spectral absorbance in terms of a number density normalized linestrength $S^{*}(T)$ [\un{cm^{-1}/molecule-cm^{-2}}]. Here, absorbance is proportional to the absorbing species number density $n_i$ [\un{molecule/cm^3}] which is related to $X_i$ by Eq.~(\ref{eq: ni}).
\begin{equation}\label{eq: ni}
    n_i = \frac{P}{kT}X_i
\end{equation}
In both conventions, $j$ is a molecular transition with linecenter frequency $\nu\sub{o}$ [\un{cm^{-1}}], and $\phi_j(\nu-\nu\sub{o},\Delta\nu\sub{c},\Delta\nu\sub{d})$ or simply $\phi_j(\nu)$ \un{[cm]} is the lineshape function. $\Delta\nu\sub{c}$ [\un{cm^{-1}}] and $\Delta\nu\sub{d}$ [\un{cm^{-1}}] are the collisional and Doppler linewidths (FWHM) of the transition. In general, all of the terms in the integrand of Eq. (\ref{eq: alpha}) are a function of position ($\ell$) along the line of sight of the laser path. In the following sections, the Beer-Lambert Law, lineshape functions, and linestrength functions will be discussed in greater detail to develop a better understanding of how the absorption of light is modeled and what these parameters represent. 

\section{The Beer-Lambert Law}

The Beer-Lambert Law is fundamental to laser absorption spectroscopy diagnostics. It describes the attenuation of light as it passes through an absorbing medium. This relationship can be derived by solving the wave equation in a dielectric medium (Eq. (\ref{eq: EP wave})) \cite{BYU2025Optics} assuming plane wave solutions of the electric field $\mathbf{E}(\mathbf{x},t) = \mathbf{E}\sub{0}e^{j(\mathbf{k\cdot x} - \omega t)}$, and macroscopic electric polarization $\mathbf{P}(\mathbf{x},t) = \mathbf{P}\sub{0}e^{j(\mathbf{k\cdot x} - \omega t)}$.

\begin{equation} \label{eq: EP wave}
    \nabla^2\mathbf{E} - \epsilon_0 \mu_0 \frac{\partial^2\mathbf{E}}{\partial t^2} = \mu_0 \frac{\partial ^2 \mathbf{P}}{\partial t^2}
\end{equation}

If we further assume the medium is linear we can write the constitutive relation $\mathbf{P}_0 = \epsilon_0 \chi(\omega) \mathbf{E}_0$, where $\chi(\omega)$ is the electric susceptibility and describes the response of a material to an applied electric field \cite{BYU2025Optics}. The dispersion relation in dielectric mediums can then be obtained 
\begin{equation} \label{eq: k-wavenum}
    k = \frac{\omega}{c}\sqrt{1+\chi(\omega)} = \frac{\omega}{c}\sqrt{1+\chi'(\omega)+j\chi''(\omega)}
\end{equation}
which leads to the complex index of refraction:
\begin{equation} \label{eq: complex n}
    \mathcal{N}(\omega) = n(\omega) + j n\sub{I} = \sqrt{1 + \chi(\omega)}
\end{equation}
The solution to the electric field can then be written in one dimension as:
\begin{equation} \label{eq: Beer E(x,t)}
    E(x,t) = E_0 e^{-\frac{n\sub{I} \omega}{c}x } e^{ j\left( \frac{n\omega}{c}x - \omega t  + \phi_x \right)}
\end{equation}

We can recognize Eq. (\ref{eq: Beer E(x,t)}) as the Beer-Lambert Law written in terms of the electric field. We can express  Eq. (\ref{eq: Beer E(x,t)}) in terms of intensity using the fact that $I(x,t) \propto |E(x,t)|^2$. The incident light intensity is then $I_0(t) = I(0,t) \propto E_0^2$ and the transmitted light intensity $I_t(t) = I(x,t) \propto E_0^2e^{-\beta x}$. By taking the ratio to eliminate the proportionality constant, we obtain the Beer-Lambert Law $I_t(t) = I_0(t)e^{-\beta x}$ where $\beta = 2 \cdot n\sub{I} \frac{\omega}{c}$ is the absorption coefficient per unit length. The next sections derive the lineshape of the absorption coefficient and its dependence on gas properties.

\section{Lineshape Functions}

Lineshape functions describe the spectral shape of the absorption coefficient and are normalized such that: 
\begin{equation} \label{eq: phi(nu)}
    \int^\infty_{-\infty}\phi(\nu)\mathrm{d}\nu = 1
\end{equation}
We can then interpret $\phi(\nu)\mathrm{d}\nu$ as the probability that a molecule absorbs incident radiation of frequency $\in [\nu, \nu + \mathrm{d}\nu]$. From Eq. (\ref{eq: complex n}) using a binomial series expansion we can express $n\sub{I} \approx \frac{1}{2}\chi''(\omega)$. Substituting this result into the expression for the absorption coefficient we obtain $\beta = \omega/c \cdot \chi''(\omega)$ which shows that the absorption coefficient has the same lineshape as the imaginary component of the electric susceptibility. This means that $\phi(\nu)$ and $\chi''(\omega)$ have the same lineshape. This result will be used in the following discussion on homogeneous broadening mechanisms and their corresponding lineshape function.

\subsection{Homogeneous Broadening Mechanisms}

Homogeneous broadening refers to broadening mechanisms that affect each atom or molecule equally and homogeneously. A resonant atomic/ molecular transition with frequency $\omega\sub{21} = (E_2 - E_1)/\hbar$ can be described by the classical electron oscillator (CEO) model where the oscillators resonant frequency $\omega\sub{a}$ is equated to $\omega\sub{21}$ \cite{siegman1986lasers}. The CEO model consists of a fixed nucleus surrounded by an electron cloud that can be displaced by an applied electric field $E_x(t)$. The electron cloud displacement $x(t)$ is governed by the differential equation given in Eq. (\ref{eq: CEO}).
\begin{equation}\label{eq: CEO}
    \frac{\mathrm{d}^2x(t)}{\mathrm{d}t^2} + \gamma \frac{\mathrm{d}x(t)}{\mathrm{d}t} + \omega_a^2 x(t) = -\frac{e}{m}E_x(t)
\end{equation}
where $\gamma$ is the energy decay rate (damping).

The general solution of Eq. (\ref{eq: CEO}) after the applied electric field is removed is as follows
\begin{equation}
    x(t) = x(t_0)e^{-(\gamma/2)(t-t_0) + j\omega'\sub{a}(t-t_0)}
\end{equation}
where $\omega'\sub{a}$ is the exact resonance frequency given by 
\begin{equation}
    \omega'\sub{a} = \sqrt{\omega^2\sub{a} - (\gamma/2)^2}
\end{equation}

Generally, $\omega\sub{a} >> (\gamma/2)$ so we can write $\omega'\sub{a} \approx \omega\sub{a}$ \cite{siegman1986lasers}. The energy decay rate $\gamma$ consists of the radiative energy decay rate or spontaneous emission rate $\gamma\sub{rad}$ and the non-radiative energy decay rate $\gamma\sub{nr}$. Since the response of each atom is identical (homogeneous), we can extend this result to a collection of atoms. 

To find a lineshape model for this response, from the previous section, we must find a mathematical form of $\chi(\omega)$ which further requires that we relate the applied electric field to the macroscopic electric polarization $p_x = \mathbf{P} \cdot \hat{x}$. The electric polarization is defined as the net electric dipole moment per unit volume. The electric dipole is related to the displacement of the electron cloud by $\mu(t) = -e x(t)$. If we consider a volume with electric dipole density $N$, and assume that all dipoles oscillate with an initial phase $\phi_0$ and initial magnitude $\mu_{x0}$, then the macroscopic electric polarization is simply $p_x(t_0) = N \times \mu_{x0}$. However, collisions perturb the phase of the dipole oscillators randomly (referred to as dephasing), effectively reducing their contribution to the macroscopic electric polarization to zero even if the amplitude of the oscillating dipoles is unaffected \cite{siegman1986lasers}. If we now assume that collisions induce a randomization of phase at an average dephasing rate $1/T_2$ (i.e., $1/T_2$ is the collision frequency for any one individual atom), then we can write for the number of oscillators that have not suffered a collision ($N(t)$) the following differential equation
\begin{equation}
    \frac{\mathrm{d} N(t)}{\mathrm{d}t} = -\frac{N(t)}{T_2}
\end{equation}
The solution is an exponential decay in the number of coherent oscillating dipoles over time. In this case, the macroscopic electric polarization is given by
\begin{align}
    p_x(t) &= N_0 e^{-(t-t_0)/T_2} \times \mu_{x0} e^{-(\gamma/2)(t-t_0) + j\omega\sub{a}(t-t_0) + j\phi_0} \\
    &= p_{x0} e^{-(\gamma/2 + 1/T_2)(t-t_0) + j\omega\sub{a}(t-t_0) + j\phi_0} \nonumber
\end{align}
Note that the pure dephasing rate $1/T_2$ gets added to the amplitude decay rate $\gamma/2$ which includes the effects of radiative and non-radiative energy decay  (i.e., $\gamma = \gamma\sub{rad} + \gamma\sub{nr}$). This solution corresponds to a second-order linear differential equation of the form
\begin{equation}
    \frac{\mathrm{d}^2p_x(t)}{\mathrm{d}t^2} + \left( \gamma + \frac{2}{T_2} \right) \frac{\mathrm{d}p_x(t)}{\mathrm{d}t} + \omega^2\sub{a}p_x(t) = \frac{Ne^2}{m}E_x(t)
\end{equation}

We now assume plane wave solutions for $p_x(t) = \tilde{P}_xe^{j\omega t}$ and $E_x(t) = \tilde{E}_xe^{j\omega t}$ to obtain $\chi(\omega)$. The result is 
\begin{equation}
    \chi(\omega) = \frac{P_x}{\epsilon_0 E_x} = \frac{Ne^2}{\epsilon_0 m} \frac{1}{\omega^2\sub{a} - \omega^2 + j\omega \Delta\omega\sub{a}}
\end{equation}
where $\Delta\omega\sub{a} = \gamma + 2/T_2$ is the linewidth (FWHM) of the resonant transition. We can simplify this expression if we consider only frequencies near resonance so that we can write $\omega^2\sub{a} - \omega^2 \approx 2\omega\sub{a}(\omega - \omega\sub{a})$. If we now isolate the imaginary component of $\chi(\omega)$ and normalize to unity, we obtain the Lorentzian lineshape $\phi\sub{L}(\omega)$

\begin{equation}\label{eq: phiL}
    \phi\sub{L}(\omega - \omega\sub{a},\Delta\omega\sub{a}) = \frac{1}{2\pi} \frac{\Delta\omega\sub{a}}{(\omega - \omega\sub{a})^2 + (\frac{\Delta\omega\sub{a}}{2})^2}
\end{equation}
Note that $\phi\sub{L}(\omega)$ describes the lineshape due to collisions and energy decay processes. At linecenter ($\omega - \omega\sub{a} = 0$), the peak amplitude of the Lorentzian lineshape $\phi\sub{L}(\omega\sub{a})$ is given by

\begin{equation}\label{eq: phiL peak}
    \phi\sub{L}(\omega\sub{a}) = \frac{2}{\pi\Delta\omega\sub{a}}
\end{equation}
Also note that we can directly replace $\omega$ with $\nu$ to give $\phi\sub{L}(\nu - \nu\sub{o},\Delta\nu\sub{L})$ since the two frequencies are proportional to each other and the lineshape is normalized. For a more rigorous discussion of the theory presented in this section, see chapter 2 of \citeauthor{siegman1986lasers} \cite{siegman1986lasers}.

\subsection{More on Homogeneous Broadening}

In the previous section, we saw that homogeneous broadening leads to a Lorentzian lineshape. In fact, Lorentzian lineshapes generally arise from any physical process that alters the lifetime of a transition, as we will show. This is a direct consequence of Heisenberg's uncertainty principle, given in Eq. (\ref{eq: dEdt}), which states that we cannot simultaneously know the energy levels associated with a transition and their lifetime \cite{banwell2017fundamentals}.

\begin{equation}\label{eq: dEdt}
    \delta E \delta t \geq \frac{h}{2\pi}
\end{equation}

The uncertainty in the frequency of the photon absorbed due to the uncertainty in the energy of the transition can be written as $\delta\nu \approx \delta E /hc$ \cite{banwell2017fundamentals}. Combing this expression with Heisenberg's uncertainty principle we find that an uncertainty in the lifetime of the transition results in a distribution of photon frequencies that can be absorbed.

\begin{equation} \label{eq: deltanu}
    \delta\nu \geq \frac{1}{2\pi c} \cdot \frac{1}{\delta t}
\end{equation}

We can use this result to obtain an expression for the linewidth of a transition as a function of the lower ($\tau''$) and upper ($\tau'$) energy level lifetimes \cite{hanson2016spectroscopy}. 

\begin{equation}
    \Delta\nu = \frac{1}{2\pi c}\left( \frac{1}{\tau''} + \frac{1}{\tau '} \right)
\end{equation}


The lifetime of an energy level is defined as the inverse of the energy decay rate $\tau \equiv \gamma^{-1}$. Furthermore, the radiative decay rate is exactly equivalent to the Einstein ``$A$'' coefficient for spontaneous emission, $\gamma_{ji} = A_{ji}$, where the subscripts $ji$ indicate spontaneous emission from $j \rightarrow i$ \cite{siegman1986lasers}. In real atoms, spontaneous emission can occur from energy levels $E_j$ and $E_i$ to a different energy level $E_k$ if the transition is allowed. Therefore, the linewidth due to spontaneous emission alone, often referred to as natural broadening, is given by \cite{hanson2016spectroscopy}

\begin{align}
    \Delta\nu\sub{N} &= \frac{1}{2\pi c}\left( \frac{1}{\tau_i} + \frac{1}{\tau_j} \right) \\
    &= \frac{1}{2\pi c} \left(\sum_{E_k < E_j} A_{ik} + \sum_{E_k < E_j} A_{jk} \right) \nonumber
\end{align}

Collisions also alter the lifetime of transition by slightly deforming the molecule during a collision or from a quantum mechanics perspective, due to the overlap of the wave function in which the quantum system now consists of both molecules. We saw in the previous section that the contribution to the Lorentzian linewidth due to collisions was given by twice the collision frequency per atom. The collision frequency of a single atom B with all atoms A can be expressed as \cite{MGDbook} 
\begin{equation}
    Z_B = \sum_A n_A d^2_{AB}\left(\frac{8\pi kT}{ m^*_{AB}} \right)^{1/2}
\end{equation}
Combing this result with Eq. (\ref{eq: deltanu}), the collisional linewidth $\Delta\nu\sub{c}$ (FWHM) can be written as \cite{hanson2016spectroscopy}
\begin{equation}
    \Delta\nu\sub{c} = \frac{Z_B}{\pi c}
\end{equation}
which is more commonly written as shown in Eq. (\ref{eq: collisional width}) by using the ideal gas law (Eq. (\ref{eq: ni}))
\begin{equation}\label{eq: collisional width}
    \Delta\nu\sub{c} = P\cdot2\gamma\sub{mix}(T)
\end{equation}
%
%
where $\gamma\sub{mix}(T)$ [\un{cm^{-1}/atm}] is the collisional broadening coefficient for the mixture and should not be confused with the energy decay rate ($\gamma$) from the CEO model. For a mixture, the collisional broadening coefficient is written as
\begin{equation}\label{eq: gamma(T)}
    2\gamma\sub{mix}(T) = \sum_A X_A \cdot 2 \gamma_{B-A}(T_0)\left(\frac{T\sub{0}}{T}\right)^n
\end{equation}
%
%
where $\gamma_{B-A}(T_0)$ [\un{cm^{-1}/atm}] is a collisional-partner dependent broadening parameter. Generally, the linewidth of natural broadening is much smaller than that due to collisional broadening such that the Lorentzian lineshape linewidth is treated as being equal to the collisional broadening linewidth $\Delta\nu\sub{L} \approx \Delta\nu\sub{c}$ \cite{banwell2017fundamentals, siegman1986lasers, hanson2016spectroscopy}.

\subsection{Inhomogeneous Broadening}\label{sec: Inhomo brodening}

Inhomogeneous broadening in general is caused by a random distribution of molecular velocities in gases, defects in solids, or some other kind of random distribution that affects the resonance frequency of otherwise identical atoms \cite{siegman1986lasers}. To calculate the inhomogeneous complex susceptibility $\chi(\omega)$, we must then average the homogeneous complex susceptibility $\chi\sub{h}(\omega,\omega\sub{a})$ by the fraction of atoms $g(\omega\sub{a})\mathrm{d}\omega\sub{a}$ whose resonance frequency is within [$\omega\sub{a}, \omega\sub{a} + \mathrm{d}\omega\sub{a}$] \cite{siegman1986lasers}. This leads to the following integral 
\begin{equation}
    \chi(\omega) = \int_{-\infty}^\infty \chi\sub{h}(\omega,\omega\sub{a}) g(\omega\sub{a})\mathrm{d}\omega\sub{a}
\end{equation}

Note that the integral has been extended to $-\infty$ simply for mathematical convenience \cite{siegman1986lasers}. If the random distribution is assumed to be a Gaussian distribution, then inhomogeneous broadening results in a Gaussian distribution of frequency shifted Lorentzian lineshapes.
The resulting complex integral however, must be evaluated each time an accurate lineshape is required. For this reason, a Gaussian lineshape is more generally used for any inhomogeneous broadening mechanism to avoid evaluating the integral. 

Let us now restrict our analysis to Doppler broadening, which results from molecules with velocity components in the direction of the laser beam. This velocity component causes a Doppler shift in the optical frequency of the photon. Absorption then occurs when the photon's shifted frequency matches the resonant frequency of the transition. In the lab frame, it appears as if the molecules resonance frequency has been shifted and is absorbing over a distribution of lab frame frequencies $\nu$. At equilibrium, the velocity distribution function (VDF) is given by a Maxwell distribution \cite{MGDbook}. If $v_z$ is the velocity component in the direction of the beam then the VDF in the $z$ direction is given by

\begin{equation} \label{eq: Maxwellian VDF}
    f(v_z)\mathrm{d}v_z = \sqrt{\frac{m}{2\pi kT}} \exp\left(-\frac{mv_z^2}{2kT} \right)\mathrm{d}v_z
\end{equation}

The Doppler-shifted optical frequency $\nu$ for a stationary light source is given by Eq. (\ref{eq: nu shift})
\begin{equation} \label{eq: nu shift}
    \nu = \left(1 + \frac{v_z}{c}\right)\nu\sub{o}
\end{equation}
Rearranging this equation we can obtain an expression for $v_z$.
\begin{equation} \label{eq: vz shift}
    v_z = \left(\frac{\nu - \nu\sub{o}}{\nu\sub{o}}\right) c
\end{equation}

We now ask, given a distribution for $v_z$, what is the distribution of optical frequencies $\nu$ that an atom will interact with? We can answer this by using Eq. (\ref{eq: vz shift}) to change the variable of integration in Eq. (\ref{eq: Maxwellian VDF}) from $v_z$ to $\nu$. Equation (\ref{eq: f(nu)}) is then the lineshape function due to Doppler broadening.

\begin{equation} \label{eq: f(nu)}
    f(\nu)\mathrm{d}\nu = \sqrt{\frac{mc^2}{2\pi kT \nu^2\sub{o}}} \exp\left( -\frac{mc^2}{2kT} \frac{(\nu - \nu\sub{o})^2}{\nu^2\sub{o}}\right) \mathrm{d}\nu
\end{equation}

If we now compare to a Gaussian distribution
\begin{equation}
    f(x) = \frac{1}{\sigma \sqrt{2\pi}} \exp\left(-\frac{1}{2}\frac{(x-\mu)^2}{\sigma^2} \right),
\end{equation}
we can find the standard deviation $\sigma$ from the argument of the exponential term which is related to the FWHM of the lineshape by FWHM= $\sigma\sqrt{8 \ln(2)}$. It is then straight forward to show that the Doppler linewidth FWHM ($\Delta\nu\sub{d}$) is given by
\begin{equation}\label{eq: Doppler width}
    \Delta\nu\sub{d} = \nu\sub{o}\sqrt{\frac{8kT \ln(2)}{mc^2}} = 7.1623 \times 10^{-7} \nu\sub{o} \sqrt{\frac{T}{\overline{M}}}
\end{equation}
where $\overline{M}$ is the molecular weight of the absorbing species. The Gaussian lineshape $\phi\sub{D}(\nu)$ can then be written in terms of $\Delta\nu\sub{d}$ as,
\begin{equation}
    \phi\sub{D}(\nu-\nu\sub{o},\Delta\nu\sub{d}) = \frac{2}{\Delta\nu\sub{d}} \sqrt{\frac{\mathrm{ln}(2)}{\pi}} \exp\left[ -4 \ln(2) \left(\frac{\nu - \nu\sub{o}}{\Delta\nu\sub{d}}\right)^2\right]
\end{equation}

At linecenter ($\nu - \nu\sub{o} = 0$), the peak amplitude of the Gaussian lineshape $\phi\sub{D}(\nu\sub{o})$ is given by
\begin{equation}\label{eq: phiD peak}
    \phi\sub{D}(\nu\sub{o}) = \frac{2}{\Delta\nu\sub{d}} \sqrt{\frac{\ln(2)}{\pi}}
\end{equation}
Note that for $\Delta\nu\sub{d} = \Delta\nu\sub{c}$, $\phi\sub{D}(\nu\sub{o}) / \phi\sub{L}(\nu\sub{o}) = \sqrt{\pi\mathrm{ln}(2)} \approx 1.5$. Similar derivations can be found in chapter 3 of \citeauthor{siegman1986lasers} \cite{siegman1986lasers} and in reference \cite{daily1997LIF}.

\subsection{Voigt lineshape}

The last lineshape we will discuss is the Voigt lineshape which is the most commonly used lineshape model due to its ability to account for both collisional and Doppler broadening effects \cite{humlivcek1982optimized, schreier2011optimized}. The Voigt lineshape which is given in Eq. (\ref{eq: Voigt}), is the result of the convolution of a Lorentzian lineshape with a Gaussian lineshape.

\begin{equation}\label{eq: Voigt}
    \phi(\nu-\nu\sub{o},\Delta\nu\sub{c},\Delta\nu\sub{d}) = \int_{-\infty}^{\infty} \phi\sub{D}(u)\phi\sub{L}(\nu-u)\mathrm{d}u
\end{equation}

This integral is usually expressed in terms of variables $x$ and $y$ which are a nondimensional line position, and the ratio of the Lorentzian to Gaussian linewidths \cite{schreier2011optimized}.
\begin{equation}
    x = \frac{\sqrt{\ln(2)} (\nu - \nu\sub{o})}{\Delta\nu\sub{d}/2}
\end{equation}
\begin{equation}
    y = \sqrt{\ln(2)}\frac{\Delta\nu\sub{c}}{\Delta\nu\sub{d}}
\end{equation}
The Voigt lineshape can then be rewritten as:
\begin{equation}
    \phi(\nu - \nu\sub{o},\Delta\nu\sub{c},\Delta\nu\sub{d}) = \phi\sub{D}(\nu\sub{o}) \cdot K(x,y)
\end{equation}
where $K(x,y)$ is the real part of the complex error function defined as:
\begin{equation}
    K(x,y) = \frac{y}{\pi}\int_{-\infty}^{\infty} \frac{e^{-t^2}}{(x-t)^2 + y^2} \mathrm{d}t
\end{equation}

Unlike the Lorentzian and Gaussian lineshapes, the Voigt lineshape does not have an analytical solution. However, several important characteristics can be approximated. At the transition linecenter ($\nu - \nu\sub{o} = 0$) the peak amplitude of the Voigt lineshape $\phi(\nu\sub{o})$ \cite{hanson2016spectroscopy} is given by 
\begin{equation}
    \phi(\nu\sub{o}) = \phi\sub{D}(\nu\sub{o}) \cdot e^{y^2}\left[ 1 - \mathrm{erf}(y) \right]
\end{equation}
This expression allows the peak absorbance $\alpha(\nu\sub{o})$ of an isolated transition to be computed without simulating the lineshape profile,
\begin{equation} \label{eq: alpha_peak}
    \alpha(\nu\sub{o}) = S_j(T)\phi(\nu\sub{o})PX_iL
\end{equation}

The linewidth $\Delta\nu$ (FWHM) of the Voigt lineshape can also be estimated \cite{schreier2011optimized, olivero1977empirical} to an accuracy of $\leq$ 0.02-0.1\% by 

\begin{equation} \label{eq: Voigt linewidth}
    \Delta\nu = 0.5346\Delta\nu\sub{c} + \sqrt{0.2166 \Delta\nu\sub{c}^2 + \Delta\nu\sub{d}^2}
\end{equation}

\begin{figure}[h]
\centering
\includegraphics[width=0.5\textwidth]{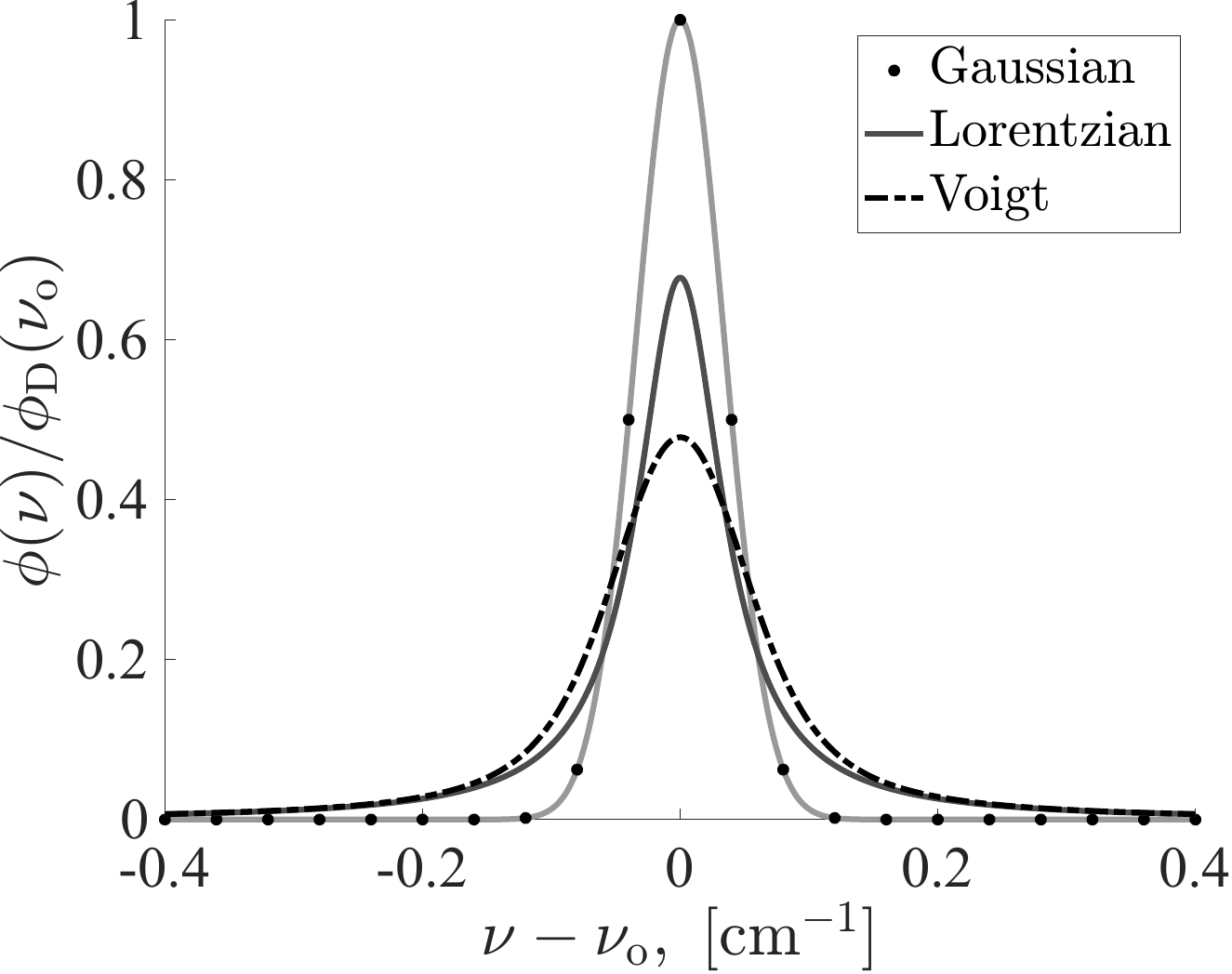}
\caption{A comparison of the Gaussian, Lorentzian and Voigt lineshape profiles with $\Delta\nu\sub{c} = \Delta\nu\sub{d}$. All lineshapes are scaled by 1/$\phi\sub{D}(\nu\sub{o})$.}
\label{fig: Lineshapes}
\end{figure}

Shown in Fig. \ref{fig: Lineshapes} is a comparison of the three lineshapes: Lorentzian $\phi\sub{L}(\nu)$, Gaussian $\phi\sub{D}(\nu)$, and Voigt $\phi(\nu)$ with $\Delta\nu\sub{c} = \Delta\nu\sub{d}$. All lineshapes have been scaled by a factor of $1/\phi\sub{D}(\nu\sub{o})$. While the Gaussian lineshape has an amplitude that is nearly 50\% greater than the Lorentzian lineshape, the amplitude in the wings approaches zero faster since all 3 lineshapes must have the same area. The Voigt lineshape has a linewidth that is larger (Eq. (\ref{eq: Voigt linewidth})) and therefore a lower peak amplitude. In the limits of $\Delta\nu\sub{c}/\Delta\nu\sub{d} << 1$ or $\Delta\nu\sub{c}/ \Delta\nu\sub{d} >> 1$, the Voigt profile converges to the lineshape that has the largest linewidth. Generally, the Doppler lineshape will be more dominant at low pressures ($P <$ 10 Torr) and the Lorentzian lineshape will be more dominant at higher pressures ($P > 1$ atm). However, the actual pressure value will depend on temperature. We can define a critical pressure ($P\sub{cr.}$) value given by Eq. (\ref{eq: Pcr})
\begin{equation} \label{eq: Pcr}
    P\sub{cr.} = \frac{\Delta\nu\sub{d}}{2\gamma\sub{mix}(T)} 
\end{equation}
where for $P\sub{cr.} << 1$ the Doppler lineshape is dominant and for $P\sub{cr.} >> 1$ the Lorentzian lineshape is dominant. A complete review of Voigt profiles and their computation is given by \citeauthor{schreier2011optimized} \cite{schreier2011optimized} and by \citeauthor{humlivcek1982optimized} \cite{humlivcek1982optimized}.
Note that there are other lineshape models not discussed including Rautian \cite{rautian1967effect} and Galatry \cite{galatry1961simultaneous} lineshapes which additionally account for Dicke narrowing \cite{dicke1953effect}, and the quadratic-speed-dependent Voigt profile (qSDVP) \cite{rohart1994speed} which accounts for speed-dependent broadening \cite{ngo2013isolated}. For a more recent discussion on these lineshapes the reader is directed to \cite{goldenstein2017infrared,goldenstein2015diode}.

\section{Transition Linestrength}

The linestrength function $S_j(T)$ of a transition $j$ is related to the state’s population $N_j$ of the absorbing species, which at equilibrium, is governed by the Boltzmann distribution given in Eq. (\ref{eq: State population}) \cite{MGDbook}
\begin{equation}\label{eq: State population}
    N_j = N \frac{C_j e^{-\epsilon_j / kT}}{Q(T)}
\end{equation}
where $Q(T)$ is the partition function and $C_j$ is the state's degeneracy. It should make sense that as the number of absorbers in a given state increases, so does the amount of light absorbed. Consequently, at a given temperature, the amount of light absorbed by a transition is proportional to the linestrength function. In the case of negligible stimulated emission, the linestrength function is given by Eq. (\ref{eq: SN}),
\begin{equation}\label{eq: SN}
    S^*_j(T) = S^*_j(T_0) \frac{Q(T_0)}{Q(T)}\exp\left[ -\frac{hc}{k}E_j'' \left(\frac{1}{T} - \frac{1}{T_0} \right)   \right]
\end{equation}
where $S^*(T_0)$ [\un{cm^{-1}/molecule-cm^{-2}}] is the linestrength at a reference temperature, $h$ [\un{J\cdot s}] is Planck's constant, $c$ [\un{cm/s}] is the speed of light, $k$ [\un{J/K}] is Boltzmann's constant, and $E''$ [\un{cm^{-1}}] is the lower-state energy of the transition. The constants ($hc/k$) are sometimes lumped into a new variable $C_2$ = 1.44 [\un{K/cm^{-1}}]. The definition given in Eq. (\ref{eq: SN}) follows the number density normalized convention; however, we can also express it following a per unit pressure basis or pressure normalized convention. In this convention, the linestrength function is given by Eq. (\ref{eq: SP}). 
\begin{equation}\label{eq: SP}
     S_j(T) = S_j(T_0) \frac{Q(T_0)}{Q(T)} \left(\frac{T_0}{T}\right) \exp\left[ -\frac{hc}{k}E_j'' \left(\frac{1}{T} - \frac{1}{T_0} \right)   \right]
\end{equation}
where $S(T_0)$ [\un{cm^{-2}/atm^{-1}}] is the reference linestrength. The conversion between the two forms is given by 
\begin{equation}
    S(T) = \frac{S^*(T) \times 0.101325 }{kT} = \frac{S^*(T) \cdot 7.34 \times 10^{21}}{T}
\end{equation}

Because it will be useful when discussing quantitative measurements in the next section, we can define the two-color ratio of linestrengths as the ratio of linestrength of two transitions (1 and 2, respectively). This ratio is given its own variable $R$ and is given by Eq. (\ref{eq: R(T)}). 
\begin{equation} \label{eq: R(T)}
   R = \frac{S_1(T)}{S_2(T)} 
    = \frac{S_1(T_0)}{S_2(T_0)}\exp\left\{\frac{hc}{k}(E_2^{\prime\prime} - E_1^{\prime\prime})\left( \frac{1}{T} - \frac{1}{T_0} 
 \right)\right\}
\end{equation}

The definition of the two-color ratio of linestrengths can be motivated by examining the ratio of the Boltzmann distribution of two transitions as expressed in Eq. (\ref{eq: Boltzmann Dist.}).
\begin{equation}\label{eq: Boltzmann Dist.}
   \frac{N_1}{N_2} = \frac{C_1}{C_2} \exp{ \left\{\frac{1}{kT}(\epsilon_2 - \epsilon_1)\right\} }
\end{equation}
We can observe that both ratios are functions of temperature, and the difference in their states energy only. Therefore, if we could measure the population of both states, we would be able infer temperature. This will be discussed in more detail the next section. 

Tabulated values for $S^*(T_0)$ and other spectroscopic parameters such as broadening parameters for many molecules of interest to combustion diagnostics with air ($\gamma_{B-\mathrm{air}}$) are available in the HITRAN online database \cite{gordon2022hitran2020}. These tabulated values can be downloaded and used with the theory presented here to simulate absorption spectra. A web-based application, \emph{SpectraPlot}, for simulating spectra of atomic and molecular transitions, which accesses the HITRAN database among others, is also available  \cite{SpectraPlot}. 


\section{Quantitative Measurements with Scanned-DA}

In this section, we describe how quantitative measurements of temperature, pressure, and species are inferred from scanned-direct absorption spectroscopy (scanned-DA) measurements of two transitions. This technique relies on tuning the laser's wavelength across an absorption feature to resolve the lineshape function and enable spectral fitting. This is done by driving the injection current with a waveform, typically a sawtooth or square wave with a leading edge, which has recently been shown to enable larger scan-depths to be achieved \cite{nair2022extended}. The spectral absorbance can be obtained from a measurement of the incident light intensity in the absence of the test gas and the transmitted light intensity with the test gas present using Beer-Lambert's law:
\begin{equation}
    \alpha(\nu) = -\mathrm{ln}\left( \frac{I\sub{t} }{I_0}\right)
\end{equation}

We now introduce the concept of the integrated absorbance which allows quantitative measurements of the thermodynamic state of the gas to be made. For an isolated transition, the integrated absorbance is given by:
\begin{equation}
    A_j = \int_{-\infty}^\infty \alpha_j(\nu) \mathrm{d}\nu 
\end{equation}
Since the lineshape function is normalized (Eq. (\ref{eq: phi(nu)})), the integrated absorbance reduces to
\begin{align} \label{eq: Aj integral}
    A_j &=  \int_0^L S_j(T)PX_i\mathrm{d}\ell \\
        &=  \int_0^L S^*_j(T)n_i\mathrm{d}\ell \nonumber
\end{align}
In the case of uniform gas properties along the line-of-sight, we can write the integrated absorbance as
\begin{align} \label{eq: Aj}
    A_j &= S_j(T)PX_iL \\
        &= S^*_j(T)n_i L \nonumber
\end{align}

\begin{figure}[h]
\centering
\includegraphics[width=0.5\textwidth]{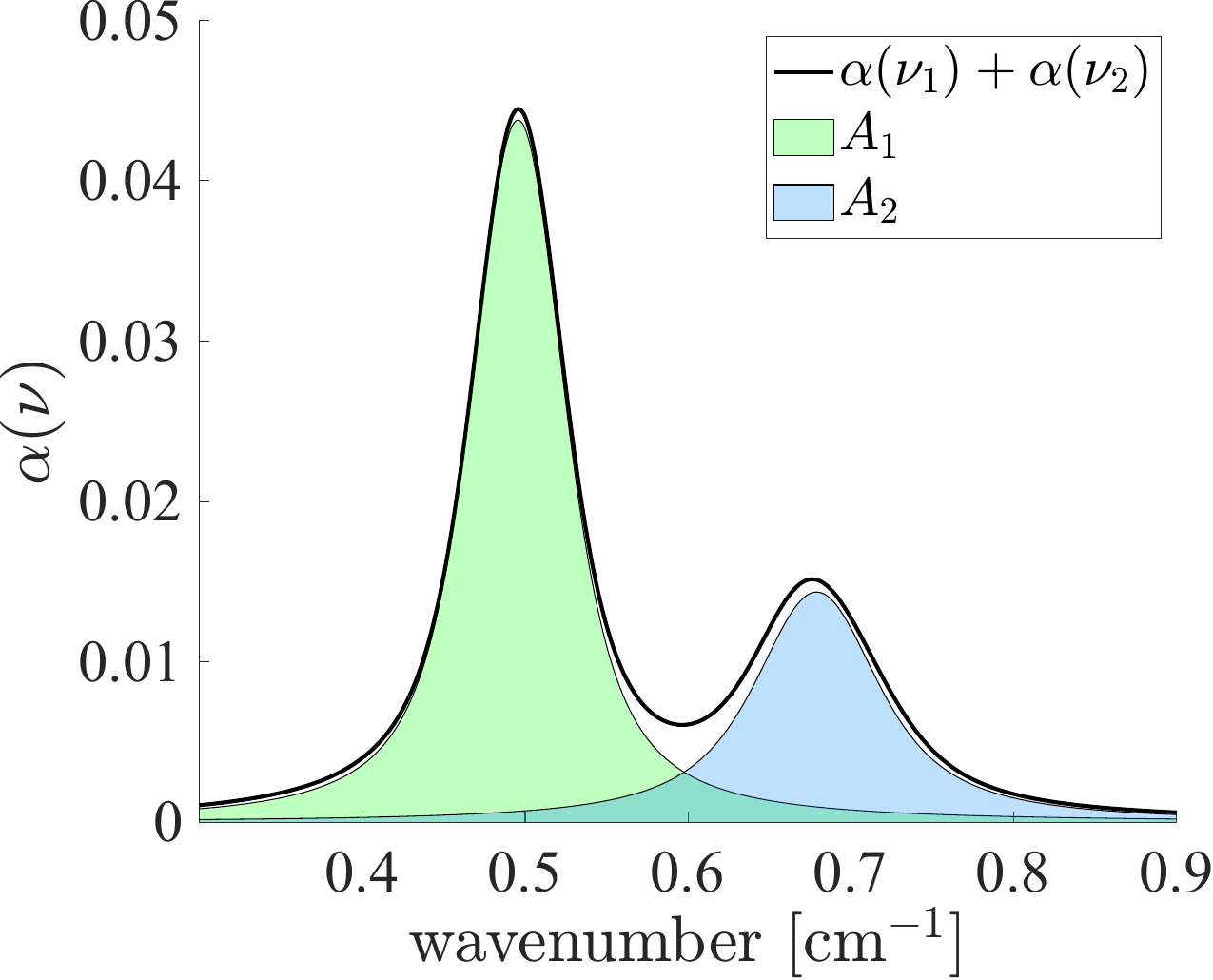}
\caption{Simulated absorption spectrum of two overlapping transitions. The integrated absorbance of each transition is shaded.}
\label{fig: VoigtArea}
\end{figure}

The above definition applies to an isolated transition. In cases where multiple closely spaced transitions overlap (line blending), the integrated absorbance is determined by reconstructing the measured absorption spectrum. If the gas properties are uniform along the laser line-of-sight, the spectral absorbance can be expressed as a linear superposition of the lineshape functions for each transition, weighted by their respective integrated absorbances:
\begin{equation}\label{eq: alpha Aj}
    \alpha(\nu) = \sum_j \alpha_j (\nu) = \sum_j A_j \phi_j(\nu-\nu\sub{o},\Delta\nu\sub{c},\Delta\nu\sub{d})
\end{equation}
This form of the spectral absorbance is useful when performing spectral fitting to infer gas properties. The integrated absorbance is never obtained through integration but is rather one of the fitting parameters \cite{goldenstein2014fitting, nair2020mhz, kuenning2024multiplexed}. Figure \ref{fig: VoigtArea} illustrates the concept of the integrated absorbance for two overlapping absorption features.

\subsection{Temperature and Partial Pressure}

Temperature and partial pressure can be obtained from the two-color (two transitions) ratio of integrated absorbances ($R$) as shown in Eq. (\ref{eq: R})
\begin{equation}\label{eq: R}
    R = \frac{A_1}{A_2} = \frac{S_1(T)}{S_2(T)} 
\end{equation}

Since the gas composition, pressure, and path length do not change for different transitions these quantities cancel out, but different molecular transitions have different temperature dependencies characterized by the linestrength and do not cancel out. This equation can then be explicitly solved for temperature as shown in Eq. (\ref{eq: T})
\begin{equation}\label{eq: T}
    T = \frac{\frac{hc}{k}\Delta E'' } {\ln(R) + \ln\left({\frac{S_2(T_0)}{S_1(T_0)}}\right) + \frac{hc}{k} \frac{\Delta E''}{T_0} }
\end{equation}
where $\Delta E'' = E_2'' - E_1''$ is the difference in lower-state energies of the two transitions. In cases where more than two lines of the same species are probed, the Boltzmann plot method can be used to infer temperature with lower uncertainties \cite{minesi2022multi, kuenning2024multiplexed}. This method relies on taking the natural logarithm of Eq. (\ref{eq: R(T)}) to write it in the form $Y_j = a E''_j + b$ where $a$ is related to temperature by $a = (hc/k)(1/T_0 - 1/T)$. With temperature now known, partial pressure be determined from the integrated absorbance of either transition as shown in Eq. (\ref{eq: Pi}).  
\begin{equation}\label{eq: Pi}
    P_i = \frac{A_j}{S_j(T)L}
\end{equation}

Notice that temperature and partial pressure can be inferred without knowledge of the gas composition or broadening parameters. If pressure is known from an external sensor, then species mole fraction can be computed as $X_i = P_i/P$. If the number density normalized convention is used instead, then the absorbing species number density can be obtained 
\begin{equation}\label{eq: ni-2}
     n_i = \frac{A_j}{S^*_j(T)L}
\end{equation}

\begin{figure}[h]
\centering
\includegraphics[width=0.55\textwidth]{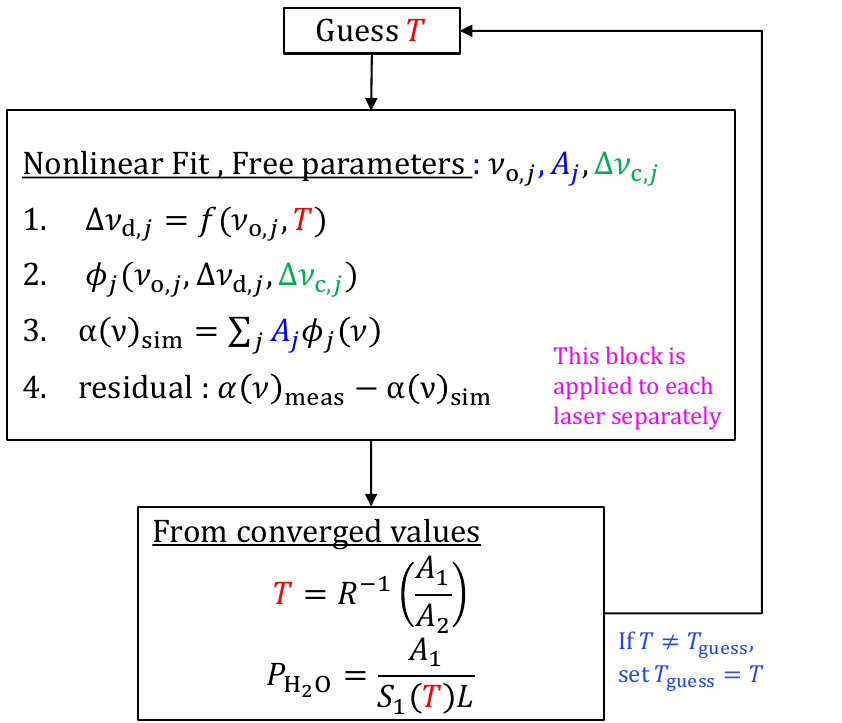}
\caption{Voigt profile fitting procedure for inferring gas properties.}
\label{fig: Voigt fitting}
\end{figure}

A general procedure for inferring gas properties using the concepts discussed thus far is provided in Fig. \ref{fig: Voigt fitting}. Note that for each transition $j$ there are generally 3 free parameters: $\nu_{\mathrm{o},j}$, $A_j$, and $\Delta\nu_{\mathrm{c},j}$. In some cases to reduce the number of free parameters, the collisional width for all transitions within the scan of each laser are assumed to be equal \cite{nair2020mhz, kuenning2024multiplexed}. This assumption should be verified for the specific lines being probed as this is not a general result. Table \ref{tab: DA err sources} provides a description of common sources that introduce errors in scanned-direct absorption spectroscopy and correction methods.

\begin{table}[p]
 \footnotesize
  \centering
  \caption{Common sources of error in scanned-direct absorption spectroscopy.}
    \begin{tabular}{p{10em}p{12em}p{18em}}
    \toprule
    \bf Source of error & \bf Effect on signal & \bf Correction method \\
    \midrule
    Reference linestrengths $S(T_0)$, and broadening parameters $\gamma_{B-A}$ & Uncertainty in $S(T_0)$ is related to the uncertainty in the inferred temperature. See Eq. (\ref{eq: T}) and section \ref{sec: UQ}. Uncertainty in $\gamma_{B-A}$ is related to the uncertainty in pressure and species mole fraction. & Measure linestrength values and broadening parameters at relevant conditions ($T$ and $P$) or take from literature when available  \cite{goldenstein2015diode, goldenstein2014wavelength, liu2007measurements}. \\[1ex]
    Baseline intensity correction & Can introduce errors in the recovered absorption spectrum if non-absorbing intensity fluctuations are not properly accounted for. & Use a non-absorbing portion in the transmitted intensity signal to scale to a reference baseline intensity signal \cite{spearrin2014simultaneous}. A non-lasing portion of the signal can also provide a second correction point to shift the intensity signal to zero. $I_\mathrm{t,cor} = (I_\mathrm{t,meas} - a)\times b$.\\[1ex]
    Pressure broadening & Can make it challenging to obtain a non-absorbing section in the transmitted intensity signal at high pressures. & Increase the scan-depth as the pressure in the absorbing environment increases and select isolated lines to reduce further line blending. \cite{nair2023optical, nair2022extended}. \\[1ex]
    Ambient absorption & Affects the recovered absorption spectrum by including absorption outside of the test gas. & Purge line-of-sight using an inert gas such as nitrogen to displace ambient air. Any non-purged region will be accounted for during the baseline intensity correction if the background is maintained steady. \\[1ex]
    Etalon effects & Constructive/destructive interference in measured intensity signals. & Tilt optics, use anti-reflection (AR) coatings, and wedged windows to avoid parallel surfaces \cite{mansfield1999evaluation}.\\[1ex]
    Thermal / wavelength drift & Causes shifts in the output wavelength over time. & Mainly a problem in fixed wavelength measurement techniques. This can be resolved by using scanned-DA to scan over a range of wavelengths such that the absorption peak is always within range. Etalon and baseline signals should also be acquired prior to each run. This is because even if the background is maintained steady, the absorption peaks location may have shifted.\\[1ex]
    \bottomrule
    \end{tabular}
  \label{tab: DA err sources}
\end{table}%

\subsection{Pressure}

Optical pressure diagnostics rely on collisional line broadening to infer pressure using Eq. (\ref{eq: collisional width}) rewritten as
\begin{equation}
    P = \frac{\Delta\nu\sub{c}}{2\gamma\sub{mix}}
\end{equation}
where $2\gamma\sub{mix}$ is the collisional line broadening coefficient which we have seen is a function of both gas composition and temperature. 

\begin{equation}
    2\gamma\sub{mix} = \sum_A X_A \cdot 2\gamma_{B-A}(T)
\end{equation}

This makes inferring pressure challenging when the gas composition is unknown. In some cases, the gas composition can be estimated using chemical equilibrium simulations that model the environment under interrogation.  Examples of such cases would be laminar premixed flat flame burners or the combustion products behind detonations in premixed detonation tubes. In other cases, the gas composition cannot be accurately determined such as in a rotating detonation engine in which fuel and air are introduced separately into the combustion chamber, with only milliseconds to mix before the arrival of the detonation wave. Inefficiencies in the combustion process also impact the composition of combustion products. In these cases, lines whose broadening parameters do not vary considerably over the range of expected gas compositions can be used \cite{nair2020mhz,nair2023optical}. For \ch{H2}-air combustion, Guerrero and Gamba demonstrated that assuming the flowfield is a binary mixture of \ch{H2O} and \ch{N2} introduces negligible errors in the collisional broadening coefficient and the inferred pressure  \cite{guerrero2025cfd}.

However, being able to estimate the gas composition is only half of the problem since an accurate spectroscopic database of broadening parameters with all collisional partners is still required. This further complicates measurements in reacting environments where intermediate species of combustion such as radicals are present for which broadening parameters cannot be easily measured. Since the broadening parameters are scaled by their corresponding mole fraction, in some cases they may be ignored ($X_A << 1$). Alternatively, models that estimate the broadening parameters based on scaling arguments can be used \cite{nair2020mhz,nair2023optical}. 

\subsection{Velocimetry}

\begin{figure}[h!]
\centering
\includegraphics[width=0.9\textwidth]{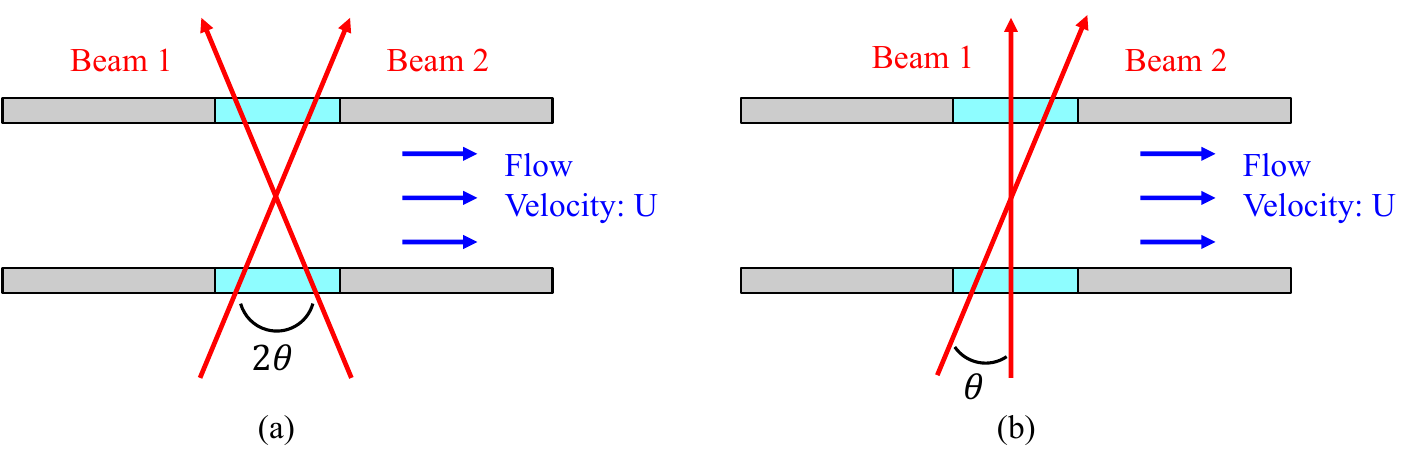}
\vspace{-0.5cm}
\caption{Typically experimental set ups for velocity measurement.}
\label{fig: velocity}
\end{figure}

Velocity measurements using TDLAS rely on application of the Doppler shift principle: 
\begin{equation}
    \nu = \left( 1 - \frac{v_z}{c}\right)\nu\sub{o}
\end{equation}
where the Doppler shift, $\Delta\nu$, in the absorbance spectrum between the two lines is measured and related to the flow velocity $U$. In practice, a single beam is split into two components that traverse the flow at an angle relative to each other and the flow direction. The frequency spacing between the linecenter of the two measured absorption spectra correspond to $\Delta\nu$. Figure \ref{fig: velocity} shows common TDLAS velocimetry experimental setups. For the configuration in Fig. \ref{fig: velocity}(a), the two beams cross at angle $2\theta$, and are oriented such that they have components in the positive and negative direction relative to the flow velocity. In this case, the flow velocity is given by:
\begin{equation}
    U = \frac{\Delta\nu }{\nu\sub{o}} \frac{c}{2\sin{(\theta)}}
\end{equation}
For the configuration in Fig. \ref{fig: velocity}(b), one beam is normal to the flow direction, while the other has a component aligned with the flow. For this setup, the flow velocity is given by:
\begin{equation}
    U = \frac{\Delta\nu }{\nu\sub{o}} \frac{c}{\sin{(\theta)}}
\end{equation}

This principle has been applied to obtain velocity measurements in various applications including the characterization of shock tubes and expansion tubes \cite{londrico2024two, schwartz2021characterization, mcgaunn2025tdlas}, direct-connect scramjet combustors \cite{schultz2014spatially}, dual-mode ramjets \cite{yun2023supersonic, yun2022spatially} and wind tunnels \cite{finch2023measurements}.

\section{Uncertainty Analysis}
\label{sec: UQ}

In this section, we present a detailed uncertainty analysis of the equations presented in the previous section. In general, for a function $y = f(x_1,x_2,...,x_N)$, the error in each variable $\Delta x_i$ (for uncorrelated variables) will contribute to the overall error in the measurement, according to Eq. (\ref{eq: unc}).
\begin{equation}\label{eq: unc}
    \Delta y = \sqrt{ \sum_i \left(  \frac{\partial f }{\partial x_i} \Delta x_i \right)^2       }
\end{equation}

For temperature measurements, errors are introduced due to the uncertainty in the reference linestrength values ($S_j(T_0)$) as well as the ratio of areas ($R$). This can be expressed as
\begin{equation} \label{eq: unc_T}
    (\Delta T)^2 = \left( \frac{\partial T}{\partial S_1(T_0)} \Delta S_1(T_0)\right)^2 + \left( \frac{\partial T}{\partial S_2(T_0)} \Delta S_2(T_0)\right)^2 + \left( \frac{\partial T}{\partial R} \Delta R\right)^2
\end{equation}
Differentiating Eq. (\ref{eq: T}) and substituting into Eq. (\ref{eq: unc_T}) we get 
\begin{equation}
    \left(\frac{\Delta T}{T}\right)^2 = \frac{ \left( \frac{\Delta S_1(T_0)}{S_1(T_0)}\right)^2 +  \left( \frac{\Delta S_2(T_0)}{S_2(T_0)}\right)^2  + \left( \frac{\Delta R}{R}\right)^2 }{ \left( \ln(R) +  \ln\left({\frac{S_2(T_0)}{S_1(T_0)}}\right) + \frac{hc}{k} \frac{E_2'' - E_1''}{T_0} \right)^2}
\end{equation}

The uncertainty in the reference linestrengths is typically reported by the HITRAN database and in literature when more accurate measurements are made. The uncertainty in $R$ can be expressed in terms of the uncertainty in the best fit areas:
\begin{equation}
    \left( \frac{\Delta R}{R} \right)^2 = \left(\frac{\Delta A_1}{A_1}\right)^2 + \left(\frac{\Delta A_2}{A_2}\right)^2
\end{equation}
$\Delta A_1$ and $\Delta A_2$ are taken to be the standard deviation ($\sigma$) in the fit parameters that are determined using Eq. (\ref{eq: Jacobian})
\begin{equation} \label{eq: Jacobian}
    \sigma =  \sqrt{ \mathrm{diag} \left(\mathrm{MSE} * (J^\dagger J)^{-1} \right) }
\end{equation}
Here, $J$ is the Jacobian matrix containing the partial derivatives of each of the fitting variables with respect to the function being fit, and MSE is the mean squared error given by \[ \mathrm{MSE} = \frac{\epsilon}{(N-p)} \] where $N$ is the number of data points, $p$ the number of fitted parameters, and $\epsilon$ is the squared L2-norm of the residual, which is simply \[ \epsilon = \left\lVert\mathrm{\alpha(\nu)_\mathrm{meas} - \alpha (\nu)_\mathrm{sim}}\right\rVert^2 \] The superscripts $\dagger$ and $-1$ indicate transpose and inverse. When using MATLAB's \emph{lsqnonlin} or \emph{lsqncurvefit} functions, $J$ and $\epsilon$ are return outputs.

We repeat this process now applying Eq. (\ref{eq: unc}) to Eq. (\ref{eq: Pi}).
\begin{equation} \label{eq: unc_H2O}
    \left( \frac{\Delta P_i}{P_i}\right)^2 = \left(\frac{\Delta A_1}{A_1} \right)^2 + \left(\frac{\Delta S_1(T)}{S_1(T)} \right)^2 + \left(\frac{\Delta L}{L} \right)^2
\end{equation}
The second term in Eq. (\ref{eq: unc_H2O}) has contributions from the uncertainty in the reference linestrength and the uncertainty in the measured temperature and can be further expressed as
\begin{equation}
    \left( \frac{\Delta S(T)}{S(T)} \right)^2 = \left(\frac{\Delta S(T_0)}{S(T_0)} \right)^2 + \left(  \frac{\partial S(T)}{\partial T} \frac{\Delta T}{S(T)}  \right)^2
\end{equation}
The partial derivative of the linestrength with respect to temperature can be numerically computed using a finite difference scheme.

\section{Line selection}

The performance of a sensor is dependent on the lines selected as well as the LAS technique used. Line selection strategies essentially reduce to choosing isolated lines with appropriate peak absorbance levels ($\alpha(\nu\sub{o})$, Eq. (\ref{eq: alpha_peak})), and lower-state energies ($E''$) \cite{goldenstein_2.5um_wavelength, zhou2005development, zhou2005selection}. In this section we briefly discuss some of the more general guidelines for sensor design.

\subsection{Lower-state Energy}

\begin{figure}[h!]
\centering
\includegraphics[width=0.9\textwidth]{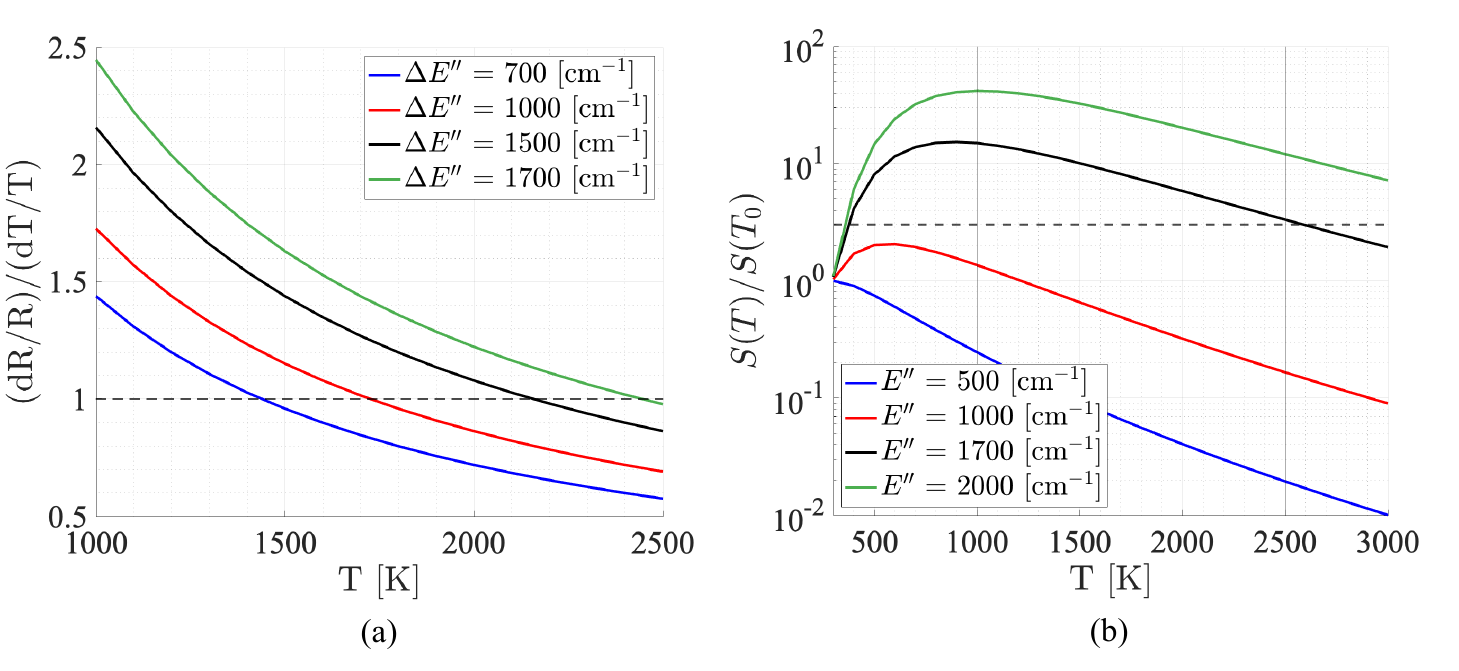}
\caption{(a) Temperature sensitivity at various $\Delta E''$ values at combustion relevant temperatures. (b) Normalized linestrength function for \ch{H2O} at various values of $E''$. The dashed line is at a normalized value of 3.}
\label{fig: dRdT}
\end{figure}

We can assess the sensitivity of a sensor (line pair) to temperature changes by differentiating Eq. (\ref{eq: R(T)}) with respect to temperature as shown in Eq. (\ref{eq: dRdT}).
\begin{equation}\label{eq: dRdT}
    \frac{ \mathrm{d} R/R}{\mathrm{d} T/T} =  \left(\frac{hc}{k}\right) \frac{\Delta E''}{T}
\end{equation}
This equation states that the percent change in the integrated absorbance ratio ($\dd R/R$) due to a percent change in temperature ($\dd T/T$) is proportional to the difference in lower-state energies of the line pair. Therefore, line pairs with a larger $\Delta E''$ are more sensitive to temperature changes. The temperature sensitivity at various values of $E''$ is shown in Fig. \ref{fig: dRdT}(a). 

In the design of high-temperature sensors, it is also desirable to minimize the influence of ambient absorption. This can be achieved by enforcing $S(T)/S(T_0)$ is at least greater than unity over the desired temperature range~\cite{zhou2005development}. This ratio is only a function of temperature and $E''$ for a given molecule. From Fig. \ref{fig: dRdT}(b), we can observe that for $E'' = 1700$ \un{cm^{-1}} the linestrength ratio is at least 3 times larger at combustion relevant temperatures (1000 \un{K} - 2500 \un{K}) than at ambient conditions. Note that this analysis should be performed with $S(T)$ and not $S^*(T)$ since with this linestrength convention absorbance is proportional to $S^*(T) n_i$ and $n_i \propto 1/T$. However, this constraint is too restricting and is generally violated for one of the lines selected.

\subsection{Non-uniformities}

\begin{figure}[h!]
\centering
\includegraphics[width=0.9\textwidth]{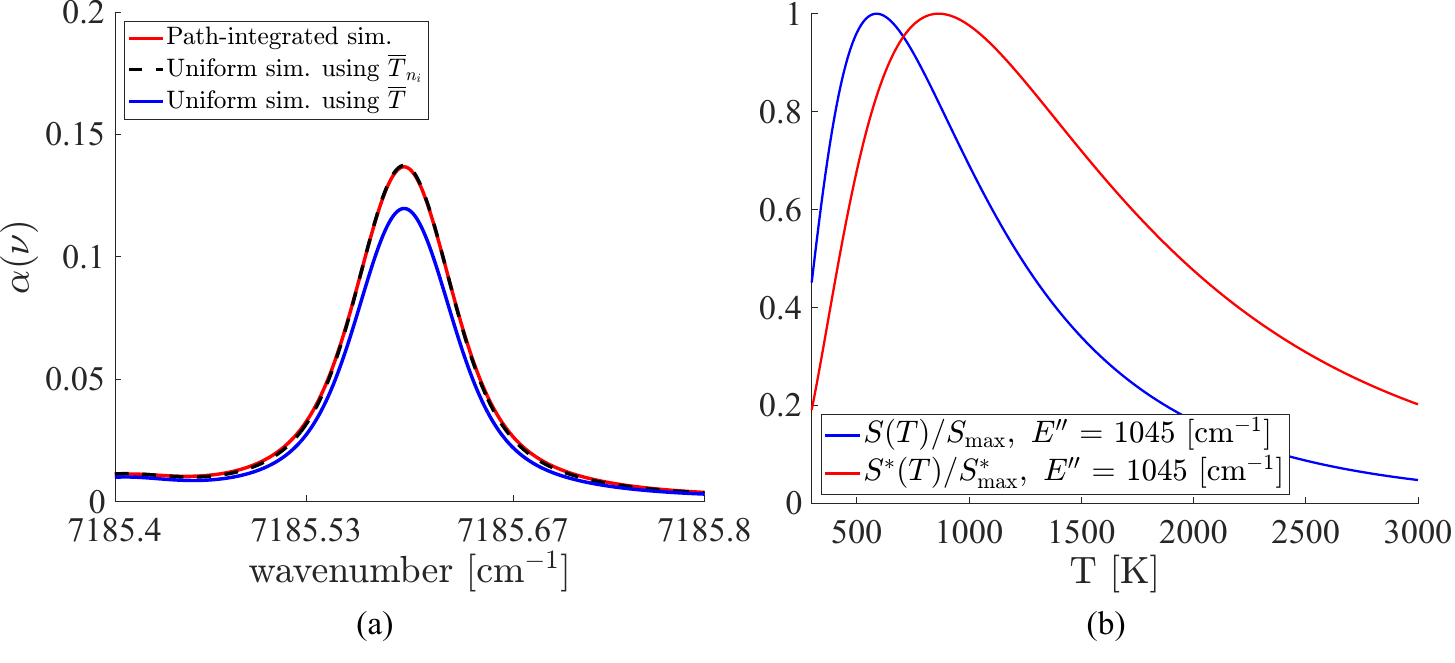}
\caption{(a) Absorption simulation of a 2-zone hypothetical temperature distribution. (b) Linestrength ratio normalized by the maximum amplitude for both unit conventions.}
\label{fig: nonuniform sim}
\end{figure}

Line selection strategies and sensors for measurements in non-uniform gases have been addressed by several researchers \cite{lin2010nonuniform_selection, liu2007nonuniform_dist, goldenstein2013nonuniform}. In LAS data post-processing, absorption spectra obtained in non-uniform flow fields is compared to simulations with uniform properties. It can be unclear what the inferred temperature represents when there is a distribution of temperatures along the line-of-sight. This is demonstrated in Fig. \ref{fig: nonuniform sim}(a) for a 2-zone temperature distribution of equal lengths with $T\sub{zone,1} = 1200$ \un{K} and $T\sub{zone,2} = 2000$ \un{K}. Pressure and species in both zones are uniform and equal. In this example, the peak absorbance is underestimated by 13\% by a simulation with uniform properties using the path-averaged temperature ($\overline{T}$). 


It has been shown analytically \cite{goldenstein2013nonuniform} that by selecting lines whose linestrengths vary linearly with temperature ($S^*(T) \approx mT + b$), the inferred temperature is equal to the number density weighted path averaged temperature ($\overline{T}_{n_i}$). 

\begin{equation}
    \overline{T}_{n_i} = \frac{\int_0^L T\cdot n_i \mathrm{d}\ell}{\int_0^L  n_i \mathrm{d}\ell}
\end{equation}

This result is significant because it allows for direct comparison of LAS measurements with CFD simulations and gives an interpretation to the inferred quantities. From Fig. \ref{fig: nonuniform sim}(a), we can see that a uniform simulation with $\overline{T}_{n_i}$ accurately predicts the peak absorbance. Species may also be compared to CFD simulations in the form of the column density $N_i$ which is given by

\begin{equation}
    N_i = \int_0^L n_i \mathrm{d}\ell = \frac{A_j}{S^{*}_j(\overline{T}_{n_i})}
\end{equation}

For transitions with pressure normalized linestrengths that vary linearly with temperature ($S(T) \approx mT + b$), the inferred temperature is equal to the partial pressure weighted path averaged temperature $\overline{T}_{P_i}$
\begin{equation}
    \overline{T}_{P_i} = \frac{\int_0^L T\cdot P_i \mathrm{d}\ell}{\int_0^L  P_i \mathrm{d}\ell}
\end{equation}
and the path-averaged partial pressure is given by
\begin{equation}
    \overline{P}_i = \frac{1}{L}\int_0^L P_i \mathrm{d}\ell = \frac{A_j}{S_j(\overline{T}_{P_i}) L}
\end{equation}
Note that the regions of linear temperature dependence differ between the two conventions because $S(T) \propto S^*(T)/T$, as shown in Fig. \ref{fig: nonuniform sim}(b). However, for practical purposes it is sufficient that the linestrength varies linearly within each individual measurement, rather than across the entire data acquisition range.

We can extend these results to linestrengths that vary quadratically with temperature: $S_j^*(T) \approx a_j T^2 + b_j T + c$. Let us first define the averaging operator $\langle \sbt \rangle _{n_i}$ to be 
\begin{equation}
    \langle \sbt \rangle _{n_i} =  \frac{\int_0^L (\sbt \cdot n_i) \mathrm{d}\ell}{\int_0^L n_i \mathrm{d}\ell}
\end{equation}
From Eq. (\ref{eq: Aj integral}), we can express the ratio of integrated absorbances as 
\begin{equation}
   \frac{A_2}{A_1} = \frac{a_2 \langle T^2 \rangle _{n_i} + b_2 \langle T \rangle _{n_i} + c_2 }{a_1 \langle T^2 \rangle _{n_i} + b_1 \langle T \rangle _{n_i} + c_1}
\end{equation}

For any averaging procedure $\langle \sbt \rangle$ and any variable $u$, we can always write $u = \langle u \rangle + u'$ where $u'$ is the deviation relative to the mean. Furthermore, for linear averaging operators, we can write $\langle u^2 \rangle  = \langle u \rangle^2 +  \langle (u')^2 \rangle$. Therefore, we can write $\langle T^2 \rangle _{n_i} = \langle T \rangle^2 _{n_i} +  \langle (T')^2 \rangle _{n_i}$.

\begin{equation}
   \frac{A_2}{A_1} = \frac{a_2 (\langle T \rangle^2 _{n_i} +  \langle (T')^2 \rangle _{n_i} ) + b_2 \langle T \rangle _{n_i} + c_2 }{a_1 (\langle T \rangle ^2 _{n_i} + \langle (T')^2 \rangle _{n_i}) + b_1 \langle T \rangle _{n_i} + c_1} = \frac{S^*_2(\overline{T}_{n_i}) + a_2\langle (T')^2 \rangle _{n_i} }{S^*_1(\overline{T}_{n_i}) + a_1 \langle (T')^2 \rangle _{n_i}}
\end{equation}

This shows that the inferred temperature is approximately equal to the number density weighted path-averaged temperature when using linestrengths that vary quadratically with temperature if the deviation component $\langle (T')^2 \rangle _{n_i}$ is negligible. 

\begin{figure}[h!]
\centering
\includegraphics[width=0.5\textwidth]{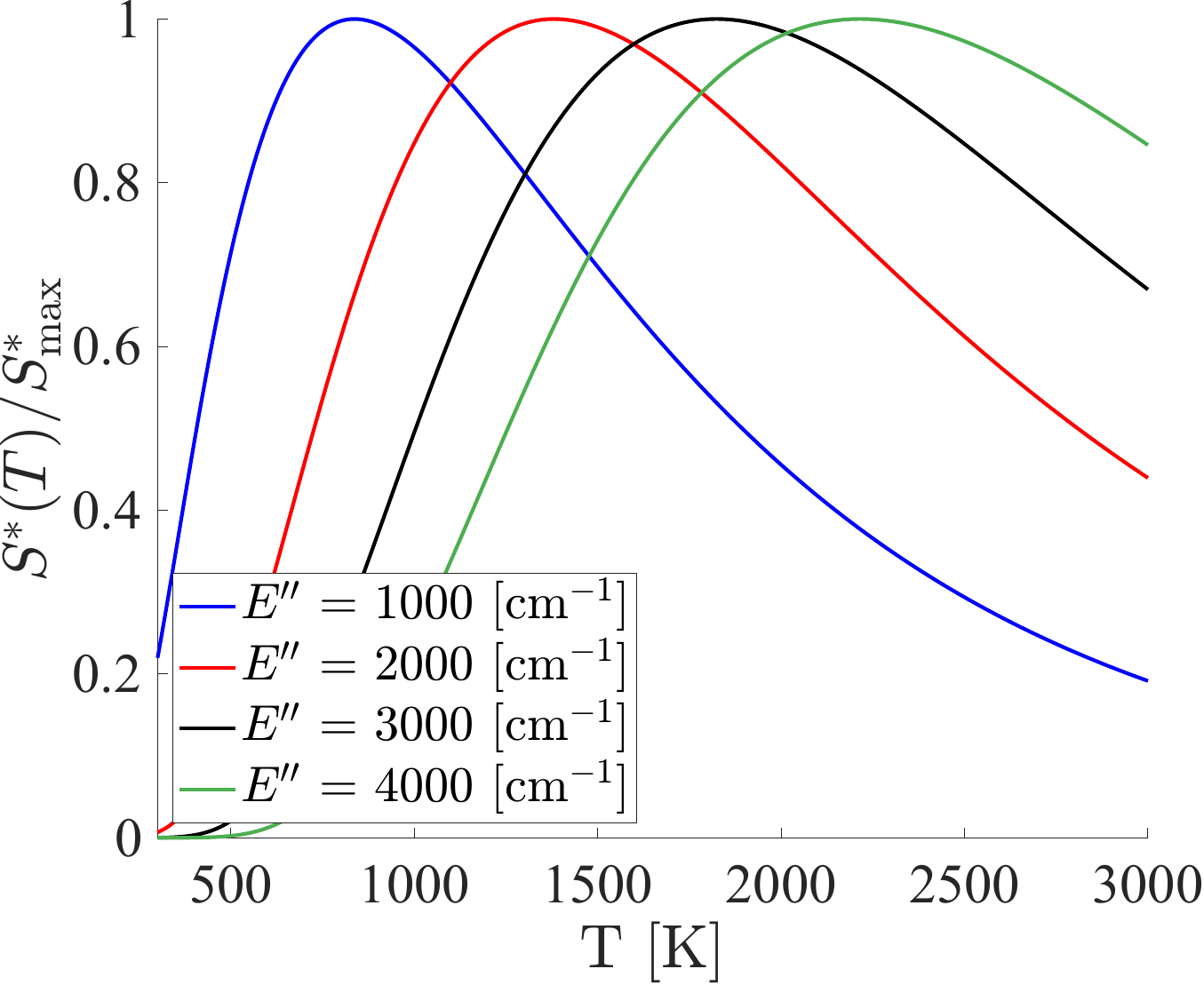}
\caption{Peak normalized linestrength versus temperature for various values of $E''$.}
\label{fig: S-Eg}
\end{figure}

While this derivation is not exact, neither is the approximation of linear linestrengths as generally the region of linear dependence will not be the same for two transitions unless the difference in their lower-state energies $\Delta E'' \lessapprox 1000$ \un{cm^{-1}}. While the assumption of linear linestrengths again only needs to hold locally,  as $\Delta E''$ increases, the overlap region where both transitions have linear linestrengths decreases and is shifted to higher temperatures as shown in Fig. \ref{fig: S-Eg}. In this case, one of the linestrengths will vary linearly with temperature ($a_1 = 0$) and the other will vary closer to quadratically with temperature. Still, the inferred temperature can be attributed to the same physical meaning.

\newpage

{\begin{center}
    \bf Part II -- Wavelength-Modulation Spectroscopy
\end{center}}

\section{Introduction to Wavelength-Modulation Spectroscopy Techniques}

Wavelength-Modulation Spectroscopy (WMS) is a laser absorption spectroscopy (LAS) technique used for its noise-rejection capabilities, ease of multiplexing multiple lasers, and robustness against external disturbances \cite{rieker2009calibration}. In WMS experiments, the wavelength of the laser is rapidly modulated about a central wavelength  while being directed across an absorbing volume of gas. Absorption information gets encoded into the harmonics of the modulation frequency $(1f\sub{mod},~ 2f\sub{mod}, ~ 3f\sub{mod}, ~ \mathrm{etc})$ which are then extracted using digital lock-in amplifiers that reject noise outside the filters passband. This WMS technique is referred to as fixed-WMS (FWMS). The more common technique however is scanned-WMS (SWMS) where the central wavelength is also simultaneously tuned across the absorption feature at a frequency $f\sub{scan} << f\sub{mod}$. This compensates for thermal drift which can shift the lasers central wavelength off the transition linecenter.

Different variations of SWMS exist and include full-spectrum SWMS, and peak-picking SWMS \cite{goldenstein2014high}. In full-spectrum SWMS the scan depth is sufficiently large to scan the central wavelength across the entire absorption feature. The WMS-$nf/1f$ harmonic signals are then extracted in the time domain using digital lock-in amplifiers \cite{sun2013analysis} or in the frequency domain using boxcar filters as outlined in \cite{schwartz2023near}. The extracted WMS signals are then fit to in order to recover gas properties \cite{goldenstein2014fitting, zhou2018compact}. This requires characterization of the lasers intensity and frequency (wavelength) response to simultaneous injection current scanning and modulation \cite{sun2013analysis}. One drawback of this method is that the sensors bandwidth using this technique can be limited by the scan depth. This is because scanning introduces numerous side bands centered at $nf\sub{mod} \pm i f\sub{scan},~i = 1, 2, 3, \mathrm{etc.}$ which when scanning over an entire absorption feature can occupy most of the frequency spectrum and prevent frequency multiplexing of a second laser \cite{goldenstein2014high}. This is overcome by decreasing $f\sub{scan}$ to make more of the frequency spectrum available.

Recent improvements to scanned-WMS-$nf/1f$ techniques have also been developed \cite{yang2019wavelength, zhu2021second,upadhyay2018new} which rely on detecting the phase-angle $\theta_{nf} = \tan^{-1}(Y_{nf}/X_{nf})$ of a single harmonic to infer gas properties. This technique has then been abbreviated as WMS-$\theta_{nf}$. Similar to WMS-$nf/1f$, WMS-$\theta_{nf}$ techniques are insensitive to the incident power on the detector and non-absorbing transmission losses \cite{peng2020analysis}. One advantage of WMS-$\theta_{nf}$ is that nonlinearities associated with frequency dependent gain ($G_{nf}$) of the detector are also canceled out. In WMS-$nf/1f$ techniques it is usually assumed that $G_{nf}$ is constant when using non-amplified biased detectors. The advantages of WMS-$\theta_{nf}$ over WMS-$nf/1f$ techniques have been quantified by analyzing the Allan deviation of WMS-$2f/1f$ and WMS-$\theta_{1f}$ signals, showing that the WMS-$\theta_{1f}$ signal is more stable over time \cite{yang2019wavelength, peng2020analysis}. Additionally, it has been observed that unlike the WMS-$nf/1f$ signals which are maximum at some modulation index greater than 1 \cite{goldenstein2014fitting}, the WMS-$\theta_{nf}$ signal peaks as the modulation index approaches zero \cite{yang2019wavelength, peng2020analysis}. This requires new strategies for the optimization of WMS-$\theta_{nf}$ sensors \cite{peng2020analysis}. 

Peak-picking SWMS refers to SWMS where the scan depth is only large enough to scan the central wavelength over the peak of the absorption feature. In this technique, the scan depth is typically around 0.01 - 0.05 [\un{cm^{-1}}] or near the transition HWHM. Due to not scanning over the entire absorption feature, a second laser can be easily multiplexed. Additionally, narrower bandwidth filters may be used which improves the signal-to-noise ratio (SNR) \cite{goldenstein2014high}. Unlike full-spectrum SWMS where the harmonic signals are fit to, in peak-picking SWMS only the amplitude at the transition linecenter is compared to simulated FWMS signals.

The most common of the two SWMS techniques is peak-picking SWMS, specifically scanned-WMS-$2f/1f$ where linecenter values are compared with FWMS signals to infer gas properties \cite{rieker2009calibration, schwartz2023near}. In this technique, the second harmonic is normalized by the first to correct for non-absorbing intensity losses such as reflections at interfaces, scattering, vibrations that move the laser beam partially off the detector, and beam steering due to density gradients \cite{rieker2009calibration,wei2019demonstration}. Since intensity losses are common to both harmonics they cancel each other out \cite{rieker2009calibration,rieker2007diode}. Normalization by the $1f$ harmonic also eliminates the dependence of extracted WMS signals on both the detector gain and incident power, enabling direct comparisons between experimental and simulated signals without scaling. This was first recognized by \citeauthor{cassidy1982atmospheric} \cite{cassidy1982atmospheric} in 1982 nearly 30 years before its widespread adoption! This approach when combined with a model for simulating WMS signals that uses laser-specific tuning parameters \cite{rieker2007diode, rieker2009calibration} is now known as calibration-free WMS to contrast previous WMS techniques using second-harmonic ($2f$) detection which did require calibration of measurements to a known condition \cite{reid1981second, philippe1993laser, liu2004wavelength, rieker2005diode}. The next section describes in more detail the standard calibration-free WMS model used for simulating FWMS and SWMS signals.

\section{Calibration-Free WMS} 
\label{sec: calibration-free WMS}

In scanned-WMS-$2f/1f$ techniques, linecenter values are compared with FWMS signals simulated using the calibration-free WMS model developed by \citeauthor{rieker2009calibration} \cite{rieker2009calibration}, which builds on the previous work of several researchers \cite{reid1981second, silver1992frequency, philippe1993laser, kluczynski2001wavelength, liu2004wavelength, li2006extension}. This model is based on the work of \citeauthor{li2006extension} \cite{li2006extension} who first introduced laser-specific tuning parameters that enable direct comparison of measurements with simulations. \citeauthor{rieker2009calibration} extended the model to be valid for any optical depth, and provided a method for $1f$ intensity normalization with background subtraction. This approach of modeling WMS signals using laser-specific tuning parameters and 1$f$ intensity normalization with background subtraction is what is referred to as calibration-free WMS. The model uses a second-order characterization of the lasers intensity (Eq. (\ref{eq: Intensity})) and a first-order characterization of the lasers frequency (Eq. (\ref{eq: wavenumber})) to describe the response of the laser to injection current modulation at a frequency $\omega = 2\pi f\sub{mod}$. 
\begin{equation} \label{eq: Intensity}
    I_0(t) = \overline{I}_0\left[ 1 + i_0 \cos(\omega t + \psi_1) + i_2 \cos(2\omega t + \psi_2) \right]
\end{equation}
\begin{equation}  \label{eq: wavenumber}
    \nu(t) = \nu\sub{o} + a \cos(\omega t)
\end{equation}

In Eqs. (\ref{eq: Intensity}) and (\ref{eq: wavenumber}), $i_0$ and $i_2$ are the first and second-order DC-normalized intensity modulation amplitudes, $\psi_1$ and $\psi_2$ are the first and second-order relative phase-shifts, $a$ [\un{cm^{-1}}] is the modulation depth and $\nu\sub{o}$ [\un{cm^{-1}}] is the transition linecenter. Note that this model does not consider scanning. The parameters $i_0$ and $i_2$ need to be measured at the transition linecenter separately for the up and down portions of the scan \cite{rieker2009calibration, mathews2020near}. This is discussed further in a later section on laser characterization. The goal of this model is to obtain analytical expressions for the WMS harmonics extracted from the detector signal using digital lock-in amplifiers. We start by first expressing the transmittance as a Fourier series expansion:

\begin{equation} \label{eq: tau}
    \tau(\nu(t)) = \left(\frac{I\sub{t}}{I_0}\right)_\nu = \exp\left\{-\alpha(\nu(t)) \right\}
\end{equation}

\begin{equation}\label{eq: tau FS}
    \tau(\nu\sub{o} + a \cos(\omega t)) = \sum_{k=0}^{k=+\infty} H_k(\nu\sub{o},a) \cos(k\omega t)
\end{equation}

Note that the sine terms in the series expansion are zero because the transmittance is an even function of time (Eq. (\ref{eq: wavenumber})). The Fourier series coefficients $H_k$ are given by
\begin{equation}
    H_0(T,P_i,\nu\sub{o},a) = \frac{1}{T} \int_{T} \tau(\nu(t)) \mathrm{d}t
\end{equation}

\begin{equation}
    H_k(T,P_i,\nu\sub{o},a) = \frac{2}{T} \int_{T} \tau(\nu(t)) \cos(k\omega t) \mathrm{d}t
\end{equation}

While these equations are already in a perfectly usable form, they are often reported differently in literature to explicitly show the dependence on the laser characterization parameters. Substituting in the right hand side of Eq. (\ref{eq: tau}) for $\tau(\nu(t))$ we obtain,
\begin{equation}
    H_0(T,P_i,\nu\sub{o},a) = \frac{1}{T} \int_{T}\exp \left\{ -\sum_j S_j(T) \phi_j(\nu\sub{o} + a \cos( \omega t)) P_i L  \right\} \mathrm{d}t
\end{equation}

\begin{equation}
    H_k(T,P_i,\nu\sub{o},a) = \frac{2}{T} \int_{T} \exp \left\{ -\sum_j S_j(T) \phi_j(\nu\sub{o} + a \cos (\omega t)) P_i L  \right\} \cos( k \omega t) \mathrm{d}t
\end{equation}

The original equations and the form commonly reported in literature are obtained by changing the variable of integration to $\theta$ using the relationship $\theta = \omega t = (2\pi/T)t$.
\begin{equation}
    H_0(T,P_i,\nu\sub{o},a) = \frac{1}{2\pi} \int_{-\pi}^{\pi} \exp \left\{ -\sum_j S_j(T) \phi_j(\nu\sub{o} + a \cos\theta) P_i L  \right\} \mathrm{d}\theta
\end{equation}

\begin{equation}\label{eq: Hk theta}
    H_k(T,P_i,\nu\sub{o},a) = \frac{1}{\pi} \int_{-\pi}^{\pi} \exp \left\{ -\sum_j S_j(T) \phi_j(\nu\sub{o} + a \cos\theta) P_i L  \right\} \cos k\theta \mathrm{d}\theta
\end{equation}

During post-processing, the detector signal which is given by $I\sub{t}(t) = I\sub{0}(t)\tau(\nu(t))$, is passed through a digital lock-in amplifier using a phase-insensitive approach to extract the $X$ and $Y$ components of the WMS-$nf$-harmonics \cite{sun2013analysis}. This is achieved by multiplying the detector signal by reference cosine and sine waves and applying a low-pass filter. Our next task is then to develop analytical expressions for the $X_{nf}$ and $Y_{nf}$ signals.

For mathematical convenience we will first multiply the incident light intensity signal by the reference cosine and sine waves, then multiply by the transmittance, and finally apply the low-pass filter. This can be expressed as:
\begin{equation} \label{eq: Icos}
I_0(t)\cos(n\omega t) = \overline{I}_0\left[ 1 + i_0 \cos(\omega t + \psi_1) +  i_2 \cos(2\omega t + \psi_2) \right] \cos(n\omega t)
\end{equation}
We now make use of the following trigonometric identity,
\begin{equation}\label{eq: trig cosx}
    \cos(\alpha)\cos(\beta) = \frac{1}{2}\left[ \cos(\beta-\alpha) + \cos(\alpha + \beta) \right]
\end{equation}
to rewrite Eq. (\ref{eq: Icos}) as
\begin{equation}
    \begin{split}
    I_0(t)\cos(n\omega t) = \overline{I}_0 \Bigl[ \cos(n\omega t) + \frac{i_0}{2}[ \cos((n-1)\omega t - \psi_1) + \cos((n+1)\omega t + \psi_1)]  + \\
    \frac{i_2}{2}[ \cos((n-2)\omega t - \psi_2) + \cos((n+2)\omega t + \psi_2)] \Bigr]
    \end{split}
\end{equation}
Next, we multiply by the Fourier series representation of $\tau(\nu(t))$ which was given in Eq. (\ref{eq: tau FS}) and apply the same trigonometric identity. Subsequently, to model the effects of a low-pass filter, we discard any high frequency terms. Typically, the filter bandwidth is $\approx 2\times f\sub{scan}$, so the only terms left are the DC components since $f\sub{mod} \approx (10 - 100) \times f\sub{scan}$. To avoid unnecessary algebra, we note that Eq. (\ref{eq: trig cosx}) produces DC components when $\alpha = \beta$. Therefore, the only remaining $H_k$ coefficients are for $k = n-2,~ n-1,~ n,~ n+1$, and $n+2$. This results in the following general expression for the $X_{nf}$ component
\begin{equation} \label{eq: Xnf}
\begin{split}
    X_{nf} = \frac{G_{nf}\overline{I}\sub{0}}{2} \Bigl[ H_n + \frac{i_0}{2} ( (1 + \delta_{n1})H_{|n-1|} + H_{n+1})\cos(\psi_1) + \\ 
    \frac{i_2}{2} ( (1 + \delta_{n2})H_{|n-2|} + H_{n+2})\cos(\psi_2)  \Bigr]
    \end{split}
\end{equation}
Note that the Kronecker-delta function ($\delta$) is used to cancel out the 1/2 term in Eq. (\ref{eq: trig cosx}) when $n = 1$ or $n = 2$ since a DC term appears without having to apply the identity. $G_{nf}$ is the frequency dependent gain of the detector. A similar expression can be obtained for the $Y$ component of the WMS-$nf$ harmonic. This time the following identity is used
\begin{equation}
    \cos(\alpha)\sin(\beta) = \frac{1}{2}\left[\sin(\beta-\alpha) + \sin(\alpha + \beta)\right]
\end{equation}
The final result for the $Y_{nf}$ component is given by
\begin{equation} \label{eq: Ynf}
    \begin{split}
    Y_{nf} = -\frac{G_{nf}\overline{I}\sub{0}}{2} \Bigl[\frac{i_0}{2} ( (1 + \delta_{n1})H_{|n-1|} - H_{n+1})\sin(\psi_1) + \\
    \frac{i_2}{2} ( (1 + \delta_{n2})H_{|n-2|} - H_{n+2})\sin(\psi_2) \Bigr]
    \end{split}
\end{equation}

Eq. (\ref{eq: Xnf}) and Eq. (\ref{eq: Ynf}) can now be evaluated for the $X$ and $Y$ components of the WMS-$nf$ harmonics. This results in the following equations for $n = 1,2,$ and $4$ which are the more commonly used WMS harmonic signals.
\begin{equation}\label{eq: X1f}
    X_{1f} = \frac{G_{1f}\overline{I}_0}{2}\left[H_1 + i_0\left(H_0 + \frac{H_2}{2}\right)\cos \psi_1 + \frac{i_2}{2}\left(H_1 + H_3\right)\cos \psi_2 \right]
\end{equation}

\begin{equation}\label{eq: Y1f}
    Y_{1f} = -\frac{G_{1f}\overline{I}_0}{2}\left[i_0\left(H_0 - \frac{H_2}{2}\right)\sin \psi_1 + \frac{i_2}{2}\left(H_1 - H_3\right)\sin \psi_2 \right]
\end{equation}

\begin{equation}\label{eq: X2f}
    X_{2f} = \frac{G_{2f}\overline{I}_0}{2}\left[H_2 + \frac{i_0}{2}\left(H_1 + H_3\right)\cos \psi_1 + i_2\left(H_0 + \frac{H_4}{2}\right)\cos \psi_2 \right]
\end{equation}

\begin{equation}\label{eq: Y2f}
    Y_{2f} = -\frac{G_{2f}\overline{I}_0}{2}\left[ \frac{i_0}{2}\left(H_1 - H_3\right)\sin \psi_1 + i_2\left(H_0 - \frac{H_4}{2}\right)\sin \psi_2 \right]
\end{equation}

\begin{equation}\label{eq: X4f}
    X_{4f} = \frac{G_{4f}\overline{I}_0}{2}\left[H_4 + \frac{i_0}{2}\left(H_3 + H_5\right)\cos \psi_1 + \frac{i_2}{2}\left(H_2 + H_6 \right)\cos \psi_2 \right]
\end{equation}

\begin{equation}\label{eq: Y4f}
    Y_{4f} = -\frac{G_{4f}\overline{I}_0}{2}\left[\frac{i_0}{2}\left(H_3 - H_5\right)\sin \psi_1 + \frac{i_2}{2}\left(H_2 - H_6 \right)\sin \psi_2 \right]
\end{equation}

Note that when no absorption is present, $H_k = \delta_{0k}$. However, the DC-normalized intensity modulation amplitude that is proportional to $\cos(n \omega t)$ in Eq. (\ref{eq: Intensity}) introduces a residual amplitude modulation (RAM) signal in the $X_{nf}$ and $Y_{nf}$ components given by:

\begin{equation}\label{eq: X1f RAM}
    X_{1f}^0 = \frac{G_{1f}\overline{I}_0}{2} i_0 \cos\psi_1
\end{equation}

\begin{equation}\label{eq: Y1f RAM}
    Y_{1f}^0 = -\frac{G_{1f}\overline{I}_0}{2} i_0 \sin \psi_1
\end{equation}

\begin{equation}\label{eq: X2f RAM}
    X_{2f}^0 = \frac{G_{2f}\overline{I}_0}{2} i_2 \cos \psi_2
\end{equation}

\begin{equation}\label{eq: Y2f RAM}
    Y_{2f}^0 = -\frac{G_{2f}\overline{I}_0}{2} i_2 \sin\psi_2
\end{equation}

\begin{equation}\label{eq: X4f RAM}
    X_{4f}^0 = Y_{4f}^0 = 0
\end{equation}

In calibration-free WMS, these RAM components need to be vector subtracted from the simulated and measured WMS signals such that the WMS-$2f/1f$ signal is zero in the absence of absorption. The simplest method for doing this is to simulate these RAM components by setting $P_{i}$ equal to zero in the absorption simulation and then extract the $X$ and $Y$ components of the WMS-$nf$ harmonics to perform background subtraction. For experimental data, a background intensity measurement obtained in the absence of absorption is used instead. The background subtracted WMS-$2f/1f$ and WMS-$4f/2f$ signals are given by Eq. (\ref{eq: bg 2f/1f}) and Eq. (\ref{eq: bg 4f/2f}).

\begin{equation}\label{eq: bg 2f/1f}
    \mathrm{WMS}-2f/1f = \sqrt{\left[ \left(\frac{X_{2f}}{R_{1f}}\right)_{\mathrm{raw}} - \left(\frac{X_{2f}}{R_{1f}}\right)_{\mathrm{bg}} \right]^2 + \left[ \left(\frac{Y_{2f}}{R_{1f}}\right)_{\mathrm{raw}} - \left(\frac{Y_{2f}}{R_{1f}}\right)_{\mathrm{bg}} \right]^2}
\end{equation}

\begin{equation}\label{eq: bg 4f/2f}
    \mathrm{WMS}-4f/2f = \frac{S_{4f}}{S_{2f}}
\end{equation}
where $R_{nf}$ and $S_{nf}$ are the absolute magnitude and absorption based magnitude of the WMS-$nf$ harmonic given by Eq. (\ref{eq: Rnf}) and Eq. (\ref{eq: Snf}).

\begin{equation} \label{eq: Rnf}
    R_{nf} = \sqrt{X_{nf}^2 + Y_{nf}^2}
\end{equation}

\begin{equation} \label{eq: Snf}
    S_{nf} = \sqrt{\left[ (X_{nf})_{\mathrm{raw}} - (X_{nf})_{\mathrm{bg}} \right]^2
    + \left[ (Y_{nf})_{\mathrm{raw}} - (Y_{nf})_{\mathrm{bg}}\right]^2 }
\end{equation}

The subscripts ``raw'' and ``bg'' refer to the intensity signal of interest obtained in an absorbing environment and the background intensity measurement respectively. Note that the background measurement includes the effects of RAM but also etalon effects and any ambient absorption that cannot be purged from the path. When simulating WMS-$nf$ signals the raw and background signals are replaced by simulations with and without absorption. Table \ref{tab: WMS error sources} lists common sources that introduce errors in the simulated WMS signals which then lead to errors in the inferred gas properties.

\begin{table}[htbp]
 \footnotesize
  \centering
  \caption{Common sources of error in calibration-free scanned-WMS.}
    \begin{tabular}{p{10em}p{12em}p{18em}}
    \toprule
    \textbf{Source of error} & \textbf{Effect on signal} & \textbf{Correction method} \\
    \midrule
    Spectral parameters & Error in modeled WMS signals & Measure linestrength values and broadening parameters at relevant conditions ($T$ and $P$) or take from literature when available.  \cite{goldenstein2015diode, goldenstein2014wavelength, liu2007measurements} \\[1ex]
    Laser tuning parameters& Error in modeled WMS signals & Verify modulation depth whenever a new set of experiments is planned and do not adjust setpoints during data collection. Intensity characterization should be performed with all optical components in place. \\[1ex]
    Ambient absorption & Affects the recovered WMS signals and inferred gas properties. & Purge line-of-sight using an inert gas such as nitrogen to displace ambient air and obtain a background intensity signal to perform background subtraction \cite{rieker2009calibration}. \\[1ex]
    Nonlinear-intensity modulation & Induces RAM components in WMS harmonic signals & Background subtraction \cite{rieker2009calibration} \\[7ex]
    Etalon effects & Constructive/destructive interference in measured intensity signals & Tilt optics, use anti-reflection coatings, and wedged windows to avoid parallel surfaces \cite{mansfield1999evaluation, mathews2021high, peng2019single, guerrero2025RDE} \\[1ex]
    Amplified detectors & The WMS harmonics are scaled by the frequency and wavelength dependent gain ($G_{nf}$) of the detector. & Measure the gain values by comparing to high-bandwidth biased detectors or determine the gain values by matching inferred gas properties to  know values \cite{mathews2020near, guerrero2025MHExT}  \\[1ex]
    Off linecenter scanning & Distorts harmonics and introduces side peaks which can be mistaken for linecenter peaks.  & Center the wavelength scan by adjusting the injection current and extracting WMS-2$f$/1$f$ values until side peaks and distortion are eliminated \cite{guerrero2025MHExT} \\[1ex]
    Thermal drift & Causes shifts in the output wavelength over time & Mainly a problem in FWMS techniques. This is mitigated by using scanned - WMS to scan a range of frequencies around the linecenter value. The WMS-$2f$ peak location is used as a linecenter reference \cite{schwartz2023near}. Acquire a background intensity signal before each run.  \\
    \bottomrule
    \end{tabular}
  \label{tab: WMS error sources}
\end{table}

\section{Derivative Spectroscopy, $a << 1$}
\label{sec: Derivative Spec}

Some insights into the shape of the Fourier series coefficients ($H_k$) can be gained if we consider the limit of small modulation depths $a<<1$ relative to the absorption features linewidth which is referred to as derivative spectroscopy \cite{reid1981second, kluczynski2001wavelength, duffin2007tunable}. If we also reduce our analysis to uniform gas properties, and optically thin samples, $\alpha(\nu) < 0.05$,  then the Beer-Lambert Law can be written as
\begin{equation}
    \tau(\nu) = 1 - \alpha(\nu)
\end{equation}
where we have made the approximation $e^x \approx 1 + x$. Since we also consider uniform gas properties we can further express the absorbance as a scaled version of the lineshape function where the scaling factor is the integrated absorbance ($A$)
\begin{equation}
    \alpha(\nu) = \big( S(T)P_iL\big) \phi(\nu) = A\phi(\nu)
\end{equation}

We can now use a Taylor series to represent the time variation in the transmittance signal instead of a Fourier series. This is done by writing a Taylor series for the lineshape function about $\nu\sub{o}$
\begin{equation}
    \phi(\nu) = \sum_{k=0}^{\infty} \frac{\phi^{(k)}(\nu\sub{0})}{k!}\Delta\nu^k
\end{equation}
where $\Delta\nu^k = (\nu - \nu\sub{o})^k = a^k\cos^k(\omega t)$ and $\phi^{(k)}(\nu\sub{o})$ is the $k$-th derivative of the lineshape function at the transition linecenter. Furthermore, we express the $\cos^k(\omega t)$ terms as functions of $\cos(m\omega t)$ terms using Eq. (\ref{eq: cosk even}) which holds for even values of $k$,
\begin{equation} \label{eq: cosk even}
    \cos^k(\omega t) = \frac{1}{2^{k-1}} \sum_{m=0}^{k/2} \binom{k}{\frac{k}{2} - m} \cos(2m \omega t)
\end{equation}
and Eq. (\ref{eq: cosk odd}) which holds for odd value of $k$,
\begin{equation} \label{eq: cosk odd}
    \cos^k(\omega t) = \frac{1}{2^{k-1}} \sum_{m=0}^{(k-1)/2}  \binom{k}{\frac{k-1}{2} - m} \cos((2m+1) \omega t)
\end{equation}
We can now make a few observations. First, $\cos^k(\omega t)$ can be written as a linear combination of $\cos(k \omega t)$ and lower frequency terms. These terms are all multiplied by $a^k / (2^{k-1} \cdot k!)$ and recall that $a << 1$. If we were to collect like terms, the largest term multiplying $\cos(\tilde{k}\omega t)$ would correspond to $k = \tilde{k}$. Therefore, we can discard all lower frequency terms by choosing $m = k/2$ for even values of $k$ and $m = (k-1)/2$ for odd values of $k$ and write a general expression for $\alpha(\nu)$

\begin{equation}\label{eq: dPhi_n}
    \alpha(\nu) \approx A \sum_{k=0}^{\infty} \frac{\phi^{(k)}(\nu\sub{0})}{2^{k}k!}a^k(2 - \delta_{k0}) \cos(k\omega t)
\end{equation}

\begin{figure}[hbt!]
\centering
\includegraphics[width=0.55\textwidth]{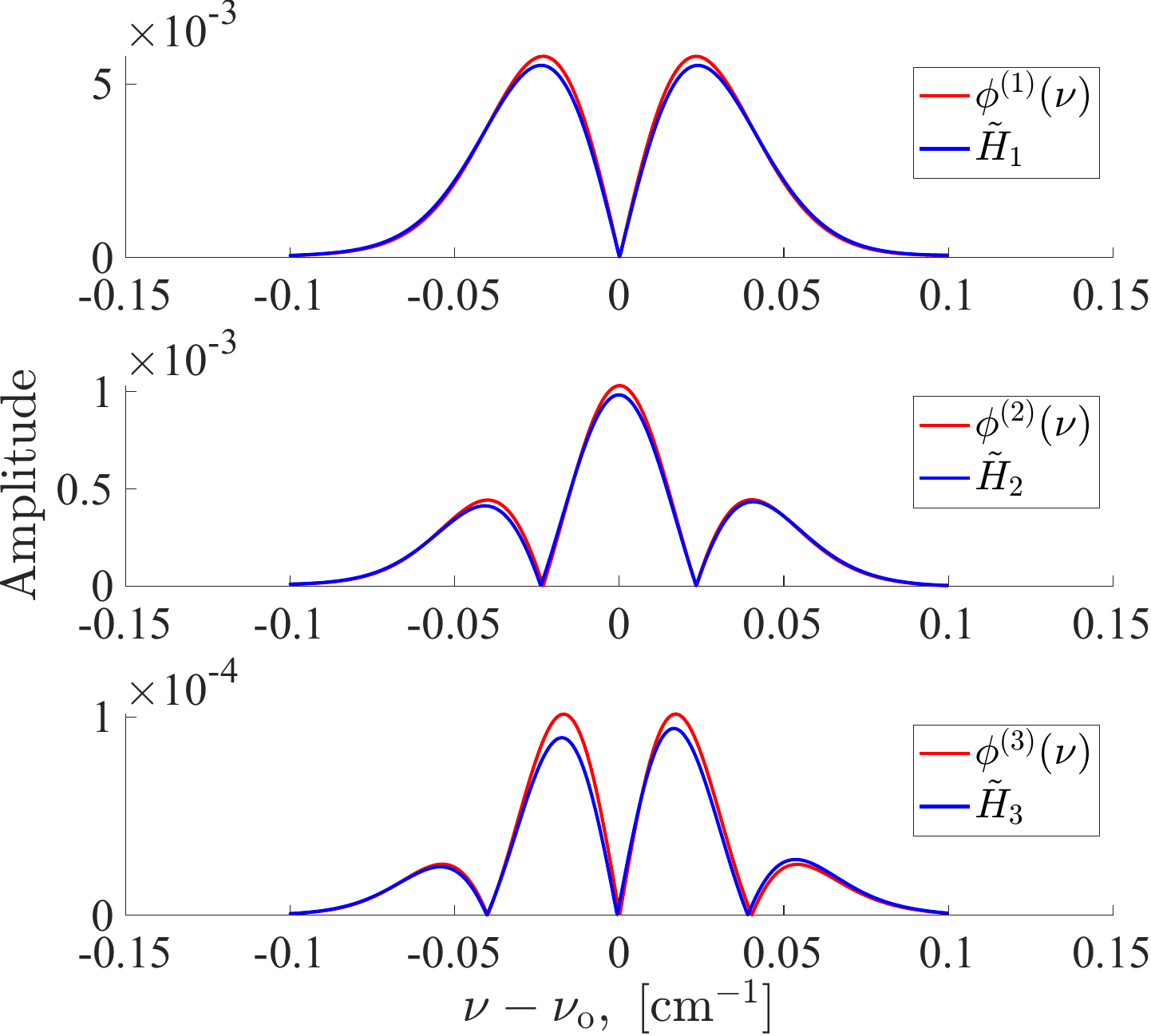}
\caption{Simulated signals of the Fourier series coefficients ($\tilde{H}_k$) and lineshape derivatives ($\phi^{(k)}$) for $a = 0.01$ \un{cm^{-1}}, $m$ = 0.35, and $\alpha(\nu\sub{o}) = 0.023.$ The lineshape derivatives are scaled by $A \cdot a^k / (2^{k-1} \cdot k!)$ as derived in Eq. (\ref{eq: dPhi_n}).}
\label{fig: derivative Hk}
\end{figure}

If we now compare the Fourier series representation of $\tau(\nu)$ in the optically thin limit \cite{li2006extension} given by
\begin{equation} \label{eq: Hk alpha}
    \tau(\nu) = 1 - \alpha(\nu) = 1- \sum_{k=0}^{\infty} \tilde{H}_k(\nu\sub{o},a) \cos(k\omega t)
\end{equation}
to the expression we just derived we can see that in the limit of small modulation depth and uniform gas properties, $\tilde{H}_k(\nu\sub{o},a)$ is proportional to the $k$-th derivative of the lineshape function. This result is shown in Fig. \ref{fig: derivative Hk} where the absolute value of the lineshape derivatives and the Fourier series coefficients are plotted for $k$ = 1, 2, and 3. Note that the Fourier series coefficients in Eq. (\ref{eq: Hk alpha}) are related to those in Eq. (\ref{eq: tau FS}) by $H_0 = 1 - \tilde{H}_0$, and $H_k = -\tilde{H}_k$ for $k > 0$ if the optically thin assumption is valid.

While this result can be used to recover absolute lineshapes \cite{duffin2007tunable}, it is of interest in noisy environments to maximize the WMS-$2f$ signal. This occurs when the modulation-index ($m$) defined by the ratio of the modulation depth to the absorption features HWHM $m = a / (\Delta\nu/2)$ is equal to 2.2 \cite{reid1981second}. For the approximation $a << 1$ to hold a modulation index of $m << 1$ is required but if higher order derivatives are kept in Eq. (\ref{eq: dPhi_n}) then larger values of $m \approx 0.3 - 0.5$ can be allowed \cite{duffin2007tunable}. This however is still far below the ideal value to maximize the WMS-$2f$ signal used in thermometry. 

\section{WMS-$2f/1f$}

In section \ref{sec: Derivative Spec} we showed that the magnitude of each sequential harmonic drops rapidly as a function of the harmonic number given by:

\begin{equation}
    \vert H_k(\nu\sub{o},a) \vert  = A\frac{ \vert \phi^{(k)}(\nu\sub{0}) \vert}{2^{k-1}k!}a^k, ~~ \mathrm{for} ~k > 0
\end{equation}

Therefore, while any harmonic normalized by the WMS-$1f$ signal is insensitive to transmission losses, the largest amplitude is given by the WMS-$2f/1f$ signal. From Eq. (\ref{eq: bg 2f/1f}) the WMS-$2f/1f$ signal assuming $\psi_1 = \pi$, and $i_2 = 0$ can be written as
\begin{equation}
\mathrm{WMS}-2f/1f = \frac{H_2 - \frac{i_0}{2} (H_1 + H_3) }{H_1 - i_0(H_0 + \frac{H_2}{2})}
\end{equation}

The WMS-$2f/1f$ amplitude at linecenter ($S_{2f/1f}(\nu\sub{o})$) in the optically thin limit ($H_k \approx \delta_{0k}$) can then be written as \cite{rieker2009calibration}
%
\begin{align} \label{eq: S_2f1f}
    S_{2f/1f}(\nu\sub{o}) &\approx \frac{H_2}{i_0} \\
    &= \frac{1}{i_0 \cdot \pi}\int_{-\pi}^{\pi} S(T) P X_i L \phi(\nu\sub{o} + a\cos \theta)\cos(2\theta)\mathrm{d}\theta
\end{align}
%
where we have made use of the fact that $H_1$ and $H_3$ are zero at the transition linecenter and $H_0 \approx 1$. This equation shows that the WMS-$2f/1f$ signal is proportional to the second derivative of the absorption lineshape and the integrated absorbance ($A = S(T)PX_iL$) if the gas properties are uniform. Proceeding with the assumption of uniform gas properties, the ratio of the WMS-$2f/1f$ signal at linecenter for two transitions becomes only a function of temperature, modulation depth ($a$), and the DC normalized intensity modulation amplitude ($i_0$). Therefore, the two-color ratio of WMS-$2f/1f$ signals can be used to infer the temperature of the gas according to
\begin{equation} \label{eq: S_2f1f Ratio}
    \frac{S_{2f/1f}(\nu\sub{o,1})}{S_{2f/1f}(\nu\sub{o,2})} = \frac{i_{0,2}}{i_{0,1}}\frac{S_1(T)}{S_2(T)}  \frac{F(\nu_{\mathrm{o},1},a_1,\phi(\nu))}{F(\nu_{\mathrm{o},2},a_2,\phi(\nu))}
\end{equation}

\begin{figure}[hbt!]
\centering
\includegraphics[width=0.9\textwidth]{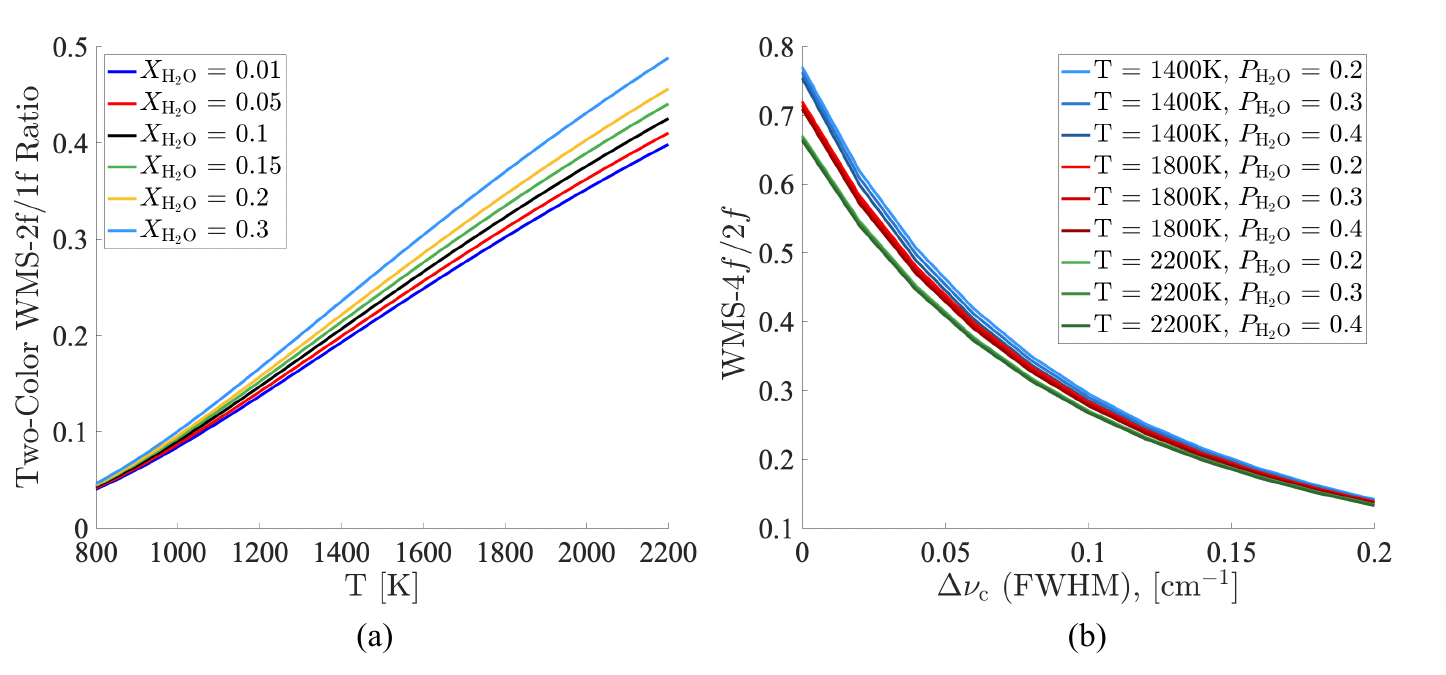}
\caption{(a) Simulation of the two-color WMS-$2f/1f$ ratio for NIR \ch{H2O} transitions. (b) Simulation of the WMS-$4f/2f$ ratio for a NIR \ch{H2O} transition plot versus the collisional width at various temperatures and partial pressures.}
\label{fig: S2f1f Ratio}
\end{figure}

Note that the ratio of integrated absorbances $S_1(T)/S_2(T) = A_1/A_2$ appears explicitly analogous to the data post-processing method for scanned-DA. While the analysis here was limited to optically thin samples and other simplifying assumptions ($\psi_1 = \pi$, and $i_2 = 0$), the model presented in section \ref{sec: calibration-free WMS} applies to any optical depth. It is also worth mentioning that $\psi_1$ increases with modulation frequency and $i_2$ increases quadratically with modulation depth \cite{li2006extension}. Therefore, Eq. (\ref{eq: S_2f1f}) should only be used at modulation frequencies less than 1 - 5 \un{kHz} and small modulation depths. In practical gas sensing when pressure is known, the WMS-$2f/1f$ signal should be simulated as a function of both temperature and the absorbing species mole fraction. Figure \ref{fig: S2f1f Ratio}(a) shows the two-color WMS-$2f/1f$ ratio simulated at a pressure of one atmosphere for NIR \ch{H2O} transitions. Although the two-color WMS-$2f/1f$ ratio is a strong function of temperature, there is a clear dependence on $X\sub{H_2O}$ which is not captured when the WMS-$2f/1f$ signals are simulated using Eq. (\ref{eq: S_2f1f}). 

\section{WMS-$4f/2f$}

In the optically thin limit ($\tau(\nu) = 1 - \alpha(\nu)$) for an isolated transition and first-order intensity characterization ($i_2 = 0$), from Eq. (\ref{eq: Hk theta}) we can write

\begin{equation}
    H_k = -\frac{S(T)PX_iL}{\pi}\int_{-\pi}^\pi \phi(\nu\sub{o} + a\cos{\theta) \cos(k\theta}) \mathrm{d}\theta, ~ ~ \mathrm{for} ~ k > 0
\end{equation}

substituting this expression into Eq. (\ref{eq: Xnf}) and Eq. (\ref{eq: Ynf}) and then substituting the result into Eq. (\ref{eq: Snf}) we find \cite{sun2013utilization}

\begin{equation}
    S_{nf}(\nu\sub{o}) = S(T)PX_iL \frac{1}{2\pi}G_{nf}\overline{I}_0 F(n,\nu\sub{o},\phi), ~ ~ \mathrm{for} ~ n \geq 2
\end{equation}

The ratio of two harmonics can then be expressed as

\begin{equation}
    \frac{S_{nf}(\nu\sub{o})}{S_{mf}(\nu\sub{o})} = \frac{F(n,\nu\sub{o},\phi)}{F(m,\nu\sub{o},\phi)}
\end{equation}

For even harmonics, $S_{nf}$ peaks at linecenter so the ratio of WMS-$4f$ peaks to WMS-$2f$ peaks is only a function of the lineshape. \citeauthor{sun2013utilization} \cite{sun2013utilization} was the first to realize this and demonstrate that the WMS-$4f/2f$ signal could be used to measure $\Delta\nu\sub{c}$ in a bath-gas of unknown composition if temperature is known. This result has been used more recently to measure species partial pressure \cite{mathews2021high, guerrero2025MHExT, Guerrero2025Quant}. Simulations of the WMS-$4f/2f$ ratio versus $\Delta\nu\sub{c}$ at various values of temperature ($\Delta\nu\sub{d}$) and partial pressure are shown in Fig. \ref{fig: S2f1f Ratio}(b). Only a weak dependence is observed with partial pressure validating that the WMS-$4f/2f$ ratio is only a function of the transition lineshape ($\phi(\Delta\nu\sub{c}, \Delta\nu\sub{d}$)).

\section{Post-Processing Algorithms for Peak-Picking SWMS}

\begin{figure}[h]
\centering
\includegraphics[width=0.9\textwidth]{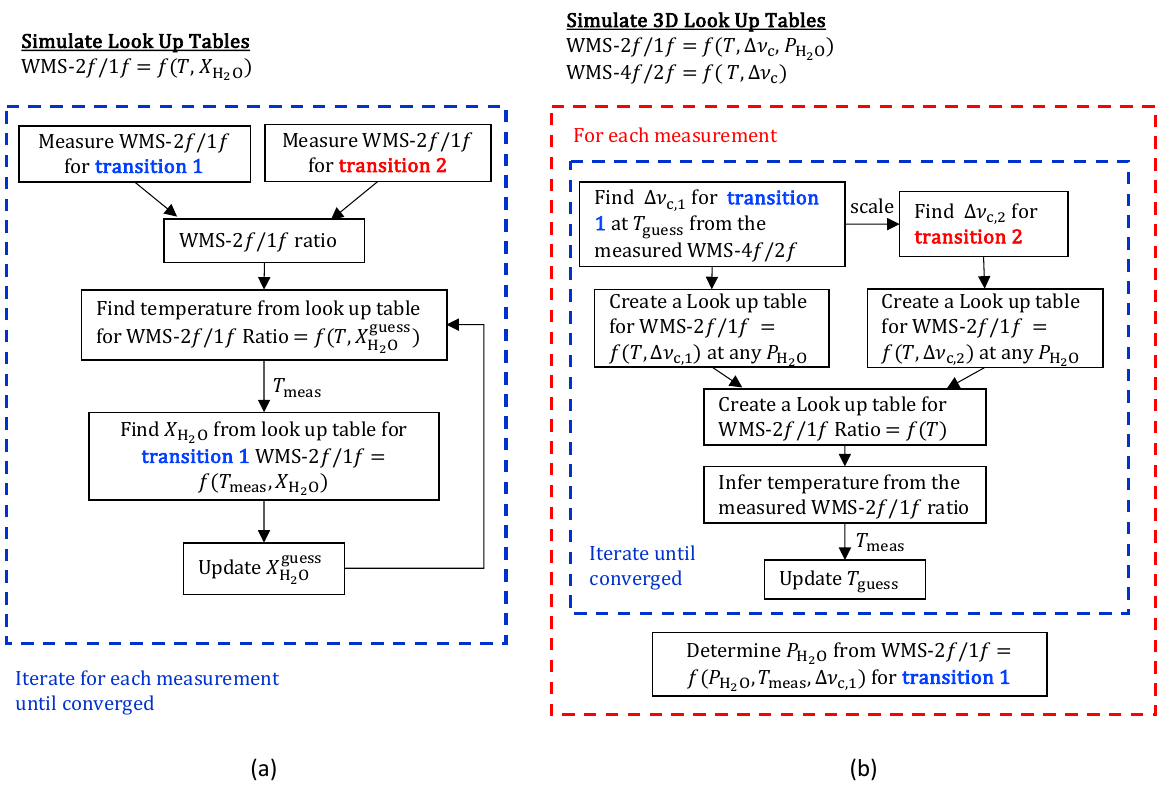}
\caption{ (a) WMS algorithms used when pressure is known. (b) WMS algorithm used when pressure is unknown and higher-order harmonics are available to extract additional spectral information.}
\label{fig: WMS-algorithms}
\end{figure}

In this section, we present two commonly used algorithms for inferring temperature and species when using scanned-WMS-$2f/1f$ techniques. The algorithms which are shown in Fig. \ref{fig: WMS-algorithms} apply to measurements acquired in environments where the pressure is known (a) and unknown (b), and are given for two generic water vapor transitions labeled 1 and 2. 

The algorithm given in Fig \ref{fig: WMS-algorithms}(a) uses look-up tables of WMS-$2f/1f$ linecenter values for both transitions simulated at the known pressure. An iterative routine is then implemented that starts by guessing a value for $X_\mathrm{H_2O}$ and looking up a value for temperature at the measured two-color WMS-$2f/1f$ ratio. The inferred temperature is subsequently used to obtain a new value for $X_\mathrm{H_2O}$ using the measured WMS-$2f/1f$ value for transition 1, which is the stronger absorbing of the two transitions. This updated value of $X_\mathrm{H_2O}$ is compared with the initial guess. If the two values differ by more than a specified tolerance, the guessed value is updated and the routine is repeated until it converges. This is done for each measurement and has been used in previous works such as \cite{goldenstein2014high,mathews2020near,schwartz2023near}.

The second algorithm given in Fig. \ref{fig: WMS-algorithms}(b) is based on the algorithm in \cite{mathews2021high} and was recently used to obtain measurements at 1 MHz in shocktubes \cite{guerrero2025MHExT}, and rotating detonation engines \cite{Guerrero2025Quant}. This post-processing algorithm applies to measurements obtained in an environment where pressure is unknown. In such cases, higher-order harmonics can be used to obtain additional spectral information. Specifically, the WMS-$4f/2f$ signal can be used to directly infer collisional width ($\Delta\nu\sub{c}$) for each transition \cite{sun2013utilization}. Additionally, because pressure is unknown, $PX_i$ is lumped together as partial pressure in Eq. (\ref{eq: alpha}), and because Eq. (\ref{eq: collisional width}) cannot be evaluated for $\Delta \nu \sub{c}$ it becomes a free parameter. Thus, absorbance is simulated as $\alpha(\nu) = \alpha(\nu,T,P_i,\Delta\nu\sub{c})$. One thing to note here is that $\Delta\nu\sub{c}$ is for the transition of interest. The collisional width of nearby transitions can be treated as being constant as done in \cite{mathews2021high} or proportional to $\Delta\nu\sub{c}$ as done in \cite{Guerrero2025Quant}. Furthermore, from Eq. (\ref{eq: S_2f1f Ratio}), we see that the two-color ratio of WMS-$2f/1f$ signals is only a function temperature and lineshape $\phi(\nu) = \phi(\nu,\Delta\nu\sub{d}(T),\Delta\nu\sub{c})$. This implies that the two-color ratio of WMS-$2f/1f$ signals is only a function of temperature and $\Delta\nu\sub{c}$ for a given linecenter frequency $\nu\sub{o}$ and laser tuning parameters. This observation is used in the post-processing algorithm.

Like the previous algorithm, look-up tables are used. Here, look-up tables of simulated WMS-$2f/1f$ linecenter values are generated for each transition, as a function of temperature, $\Delta\nu\sub{c}$, and \ch{H2O} partial pressure. Additionally, a look-up table for WMS-$4f/2f$ linecenter values is generated for the first transition as a function of temperature and $\Delta\nu\sub{c}$. 

The algorithm begins by using the measured WMS-$4f/2f$ amplitude for transition 1 to estimate $\Delta\nu\sub{c,1}$ at an initial temperature guess.  Due to MHz rate modulation frequencies, the $4f$ harmonic of the second transition was greater than the detectors bandwidth and $\Delta\nu\sub{c,2}$ was instead approximated by scaling $\Delta\nu\sub{c,1}$ \cite{mathews2021high,Guerrero2025Quant}. Once $\Delta\nu\sub{c,1}$ and $\Delta\nu\sub{c,2}$ are known, a look-up table for the two-color WMS-$2f/1f$ ratio is interpolated at the measured $\Delta\nu\sub{c}$ values and any $P_\mathrm{H_2O}$ since the two-color is independent of this choice. Temperature is then inferred from the measured two-color WMS-$2f/1f$ ratio and compared to the guessed value. If the two temperature values have not converged, the guessed temperature is updated and the process is repeated. Finally, partial pressure is inferred from the WMS-$2f/1f$ look-up table for transition 1 using the measured WMS-$2f/1f$ signal amplitude, converged temperature, and $\Delta\nu\sub{c,1}$. 

\section{Development of MHz Rate SWMS Sensors}
\label{sec: MHz rate freq}

The development of high-speed ($>100$ kHz) SWMS sensors has been limited by several factors. The focus of this section is on understanding these limitations and to highlight recent advancements that have enabled MHz rate LAS sensors (scanned-DA, and SWMS) to be developed. 

During post-processing of peak-picking SWMS data, digital lock-in amplifiers are used to extract the $X$ and $Y$ components of the harmonic signal typically using filters with a bandwidth of 2.01 - 2.2$\times f\sub{scan}$ \cite{goldenstein2014high,mathews2021high}. To avoid overlap of side bands from the harmonics of the second frequency multiplexed laser, we can impose a minimum spacing between the modulation frequencies as shown in Eq. (\ref{eq: delta-fm}).

\begin{equation}\label{eq: delta-fm}
     f\sub{mod,2} - f\sub{mod,1} > 4.4\times f\sub{scan}
\end{equation}

We can do the same for the $2f$ harmonics which would decrease the multiplicative factor on $f\sub{scan}$ by 2. Additionally, we want for the $3f$ harmonic of the laser with a lower frequency to not coincide with the $2f$ harmonic of the second laser. This constraint is given in Eq. (\ref{eq: 3f-cons}).

\begin{equation}\label{eq: 3f-cons}
    3f\sub{mod,1} - 2f\sub{mod,2} > 4.4\times f\sub{scan}
\end{equation}

\begin{figure}[h]
\centering
\includegraphics[width=0.9\textwidth]{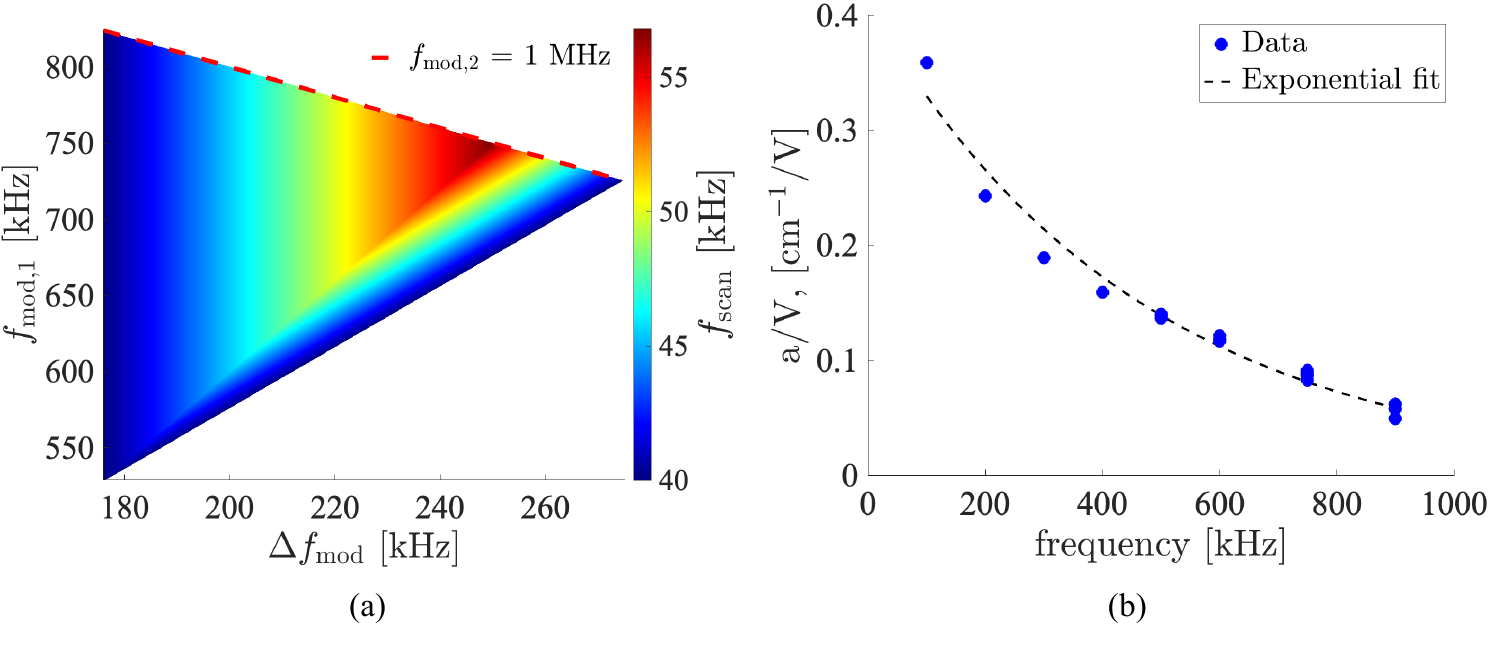}
\caption{ (a) Contour plot of the maximum $f\sub{scan}$ that satisfies the given constraints for a given $f\sub{mod,1}$ and $\Delta f\sub{mod}$. (b) Achievable modulation depth of a NIR tunable diode laser per peak-to-peak voltage of the applied sinusoidal waveform over frequency. }
\label{fig: fmod-req}
\end{figure}

A third constraint is imposed by the laser controller's bandwidth which is that $f\sub{mod,2} \leq 1~\mathrm{MHz}$. A contour map of the largest $f_{\mathrm{scan}}$ that meets these constraints is shown in Fig. \ref{fig: fmod-req}(a) where $\Delta f\sub{mod}$ is defined as $f\sub{mod,2} - f\sub{mod,1}$ and shows that for a scanning frequency of 50 kHz (measurement rate of 100 kHz), the modulation frequencies required are already near 1 MHz. Secondly, we notice that the largest possible $f_{\mathrm{scan}}$ is approximately 57 \un{kHz}. To increase $f_{\mathrm{scan}}$, it becomes necessary to use higher modulation frequencies and to increase $\Delta f\sub{mod}$ which is well known by researchers in the community.

A fourth constraint can be imposed to reduce the number of side bands which is to select modulation frequencies divisible by $f_{\mathrm{scan}}$ so that every scan is periodic, see section 3.2 of reference \cite{strand2014scanned}. This constraint for $f_{\mathrm{scan}} = 50 ~\mathrm{kHz}$ would make $(f_{\mathrm{mod,1}},f_{\mathrm{mod,2}}) = (750 ~\mathrm{kHz}, 1 ~\mathrm{MHz})$ the only pair that meets all constraints.

Another important aspect we have yet to consider is the steep decrease in modulation depth with increasing modulation frequency \cite{mathews2020near}. To maximize the WMS-$2f$ harmonic signals, it has been shown that a modulation depth $a = 2.2\Delta\nu ~(\mathrm{HWHM})$ is required \cite{reid1981second}. This constraint ultimately imposes an upper limit on $f\sub{mod}$ or limits the sensor to low pressure applications since $\Delta\nu \propto P$. Shown in Fig. \ref{fig: fmod-req}(b) is the exponential decrease in the tuning capabilities of a tunable diode laser emitting near 7185.596 \un{cm^{-1}}. 

These simple calculations illustrate why SWMS sensors have generally been limited to measurement rates of 50 kHz when modulating the injection current through the laser controller \cite{goldenstein2014high,cassady2021time}. Note that the sensor demonstrated in \cite{goldenstein2014high} violates some of the constraints given here most likely due to insufficient modulation depth at higher frequencies. In this case, narrower filters were used to extract the $1f$ harmonic than the $2f$ harmonic. This analysis also shows why generally $f\sub{mod} \approx (10-100)\times f\sub{scan}$.

By using bias-tee circuitry, MHz frequencies outside of the bandwidth of the LDC can be achieved, and equally as important it has also been shown that modulation depth recovers due to the absence of parasitic electrical components in the path \cite{mathews2020near, nair2022extended}. This fortuitous result is what has enabled MHz measurement rate scanned-WMS \cite{mathews2021infrared, mathews2020near, guerrero2025RDE, guerrero2025MHExT} and scanned-DA sensors \cite{nair2020mhz, nair2023optical} to be developed.

\section{Intensity and Wavelength Characterization}

Up to now, we have not discussed how the laser specific tuning parameters ($i_0,~i_2,$ $\psi_1,~\psi_2,~a$) which are inputs to the calibration-free WMS model are obtained. The procedure for intensity characterization of FWMS signals \cite{li2006extension} and for modulation depth at both kHz \cite{XChao2023utility} and MHz \cite{mathews2020near} rate modulation frequencies have been thoroughly documented so only a brief review is given here. However, we provide a more detailed discussion of intensity characterization for SWMS. 

\begin{figure}[htb!]
\centering
\includegraphics[width=0.9\textwidth]{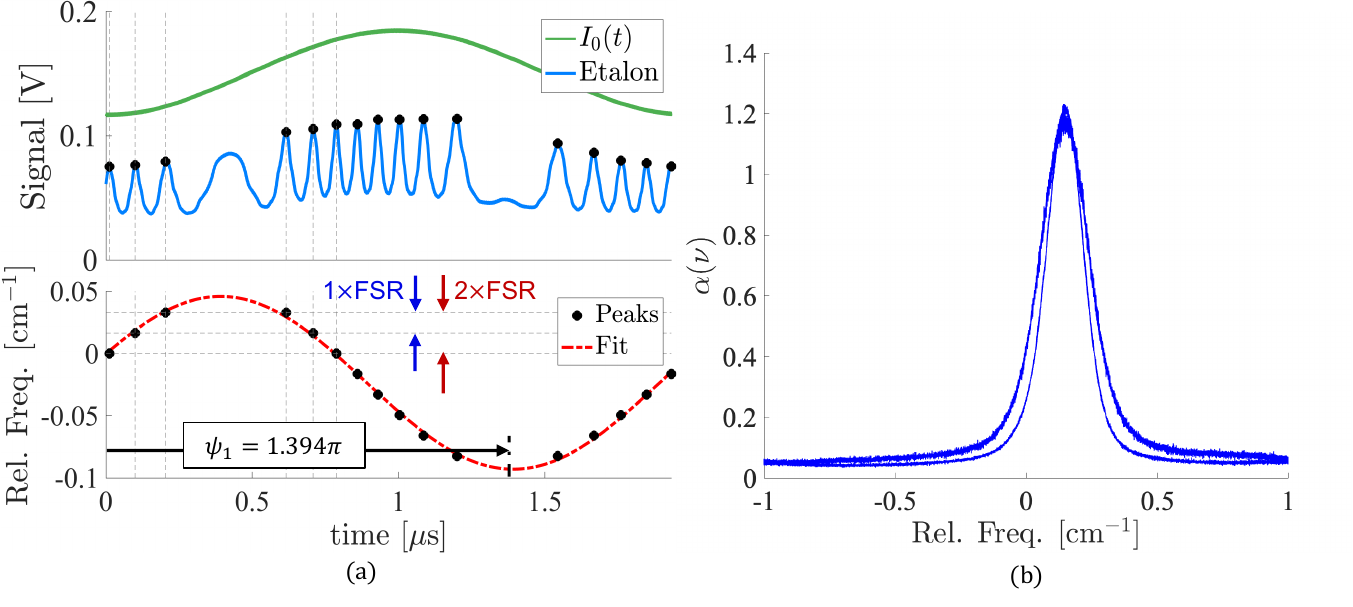}
\caption{(a) (top) Baseline and etalon intensity signals. (bottom) Identified etalon peaks evenly spaced in time by the FSR [\un{cm^{-1}}] of the etalon. (b) Absorbance plotted versus Eq. (\ref{eq: v-fit}).}
\label{fig: psi1 char}
\end{figure}

All parameters can be obtained from a baseline intensity measurement in the absence of the test gas with all optical components in place, and an etalon signal which measures the relative frequency (wavelength) change in time. Example FWMS signals are shown in Fig. \ref{fig: psi1 char}(a) for light modulated at 500 kHz. While $i_0$ and $i_2$ can change for different optical setups, $\psi_1,~\psi_2$ and $a$ are intrinsic to the laser.

We start by determining the DC normalized intensity modulation amplitudes by fitting Eq. (\ref{eq: Intensity}), to the measured baseline intensity signal in two steps by expressing $I_0(t) = f(\omega t) + g(2\omega t)$. First, $f(\omega t)$, given in Eq. (\ref{eq: i0-fit}), is fit to the measured baseline intensity signal
\begin{equation} \label{eq: i0-fit}
    f(\omega t) = I\sub{DC} + i^*_0 \cos(\omega t + \varphi_1) 
\end{equation}
and then Eq. (\ref{eq: i2-fit}) is fit to the residual
\begin{equation} \label{eq: i2-fit}
    g(2\omega t) = i^*_2 \cos(2\omega t + \varphi_2) 
\end{equation}

The DC normalized intensity modulation amplitudes are then given by $i_0 = i^*_0/I\sub{DC}$ and $i_2 = i^*_2/I\sub{DC}$. Next, we can determine the modulation depth ($a$). Here, we use the etalon signal which produces peaks that are evenly spaced in time by the free-spectral range (FSR [\un{cm^{-1}}]) unique to the etalon. Each sequential peak will be located at $\pm 1$ FSR relative to the previous or +0 FSR if the peak occurs after a frequency reversal \cite{XChao2023utility}. Once the etalon peaks are located and labeled, Eq. (\ref{eq: a-fit}) can be fit to the peaks to determine $a$. Guidelines and a MATLAB GUI for peak labeling are provided in reference \cite{XChao2023utility}.
\begin{equation}  \label{eq: a-fit}
    \nu(t) = b + a \cos(\omega t + \phi)
\end{equation}
Note that in Eq. (\ref{eq: a-fit}), $b$ is not necessarily the transition linecenter $\nu\sub{o}$. Since etalons can only provide a measurement of the relative frequency (wavelength) change in time, a reference absorption signal is necessary to determine $b$. Finally, the relative phase shift is given by $\psi_1 = \varphi_1 - \phi$ and the nonlinear phase shift by $\psi_2 = \varphi_2 - 2\phi$. 

In the case of simultaneous sinusoidal wavelength scanning (SWMS), the general intensity and frequency characterizations are given by
\begin{equation} \label{eq: Intensity scan}
    I_0(t) = \overline{I}_0\left[ 1 + i\sub{s} \cos(\omega\sub{s} t + \varphi\sub{s}) + i_0 \cos(\omega t + \varphi_1) + i_2 \cos(2\omega t + \varphi_2) \right]
\end{equation}

\begin{equation}  \label{eq: I0 scan}
    \nu(t) = b +  a\sub{s} \cos(\omega\sub{s}t + \phi\sub{s}) + a \cos(\omega t + \phi\sub{m})
\end{equation}
Since the calibration-free WMS model does not consider scanning, we need to lump the terms associated with scanning with the DC components

\begin{equation} \label{eq: wn scan}
    I_0(t) = \overline{I}\sub{DC}(t) + i^*_0 \cos(\omega t + \varphi_1) + i^*_2 \cos(2\omega t + \varphi_2)
\end{equation}

\begin{equation}  \label{eq: wavenumber scan}
    \nu(t) = \overline{b}(t) + a \cos(\omega t + \phi\sub{m})
\end{equation}
where $\overline{I}\sub{DC}(t)$ is the time varying DC intensity
\begin{equation}
    \overline{I}\sub{DC}(t) = \overline{I}_0 + i^*\sub{s} \cos(\omega\sub{s} t + \varphi\sub{s})
\end{equation}
and $\overline{b}(t)$ is the time varying linecenter frequency
\begin{equation} \label{eq: b(t)}
    \overline{b}(t) = b + a\sub{s} \cos(\omega\sub{s} t + \phi\sub{s})
\end{equation}

These equations are now in the correct form for the calibration-free WMS model but note that $i_0 = i^*_0/I\sub{DC}(t)$ is now a function of time. When comparing SWMS signals at linecenter to simulated FWMS signals, $i_0$ and $i_2$ have to be measured at the transition linecenter ($\overline{b}(t) = \nu\sub{o}$) using an absorption reference. Also note that $i_0$ is different during the up-scan and down-scan due to a different instantaneous DC offset caused by the relative phase shift $\varphi\sub{s} - \phi\sub{s}$. However, in the case of uniform properties from Eq. (\ref{eq: S_2f1f Ratio}) by setting $\nu\sub{o,1} = \nu\sub{o,2}$ their magnitudes are related through Eq. (\ref{eq: i0_peaks}).

\begin{equation}\label{eq: i0_peaks}
    \frac{S_{2f/1f}^{\mathrm{up}}(\nu\sub{o})}{S_{2f/1f}^{\mathrm{down}}(\nu\sub{o})} = \frac{i^{\mathrm{down}}_0}{i^{\mathrm{up}}_0}
\end{equation}

Also note that because in peak-picking SWMS experiments the wavelength of the laser light is tuned across the absorption feature and only the transition linecenter values are used to infer gas properties, when simulating FWMS signals using Eqs. (\ref{eq: Intensity})-(\ref{eq: wavenumber}), $\nu\sub{o}$ in Eq. (\ref{eq: wavenumber}) is set to the known transition linecenter. On the other hand, if full-spectrum SWMS is used, then $b$ in Eq. (\ref{eq: b(t)}) must be determined using an absorption reference \cite{goldenstein2014fitting, sun2013analysis}. 

For SWMS wavelength characterization, modulation depth and scan depth can be measured independently by turning off one of the waveforms, since the laser diodes response to each waveform is generally linear \cite{goldenstein2014fitting}. At modulation frequencies much above a few MHz however, an etalon is no longer the best way to characterize the modulation depth due to the increasing demand on the sampling rate required to resolve the etalon peaks between scans. Instead, a scanning Fabry-Perot Interferometer (FPI) \cite{hercher1968spherical} which scans a certain range of frequencies as the spacing between two confocal spherical mirrors is changed (via a piezoelectric transducer) is used to analyze the intensity spectrum of the light when modulated \cite{mathews2020near}. The width of the intensity spectrum is primarily dependent on the modulation depth which enables accurate measurements. \citeauthor{silver1992frequency} \cite{silver1992frequency} provides a theoretical description of frequency modulated light and \citeauthor{mathews2020near} \cite{mathews2020near} describe the procedure for using an FPI to characterize the lasers modulation depth. Here, rather than repeat the discussion which is completely thorough, we provide an overview on the operation of an FPI and the required signals for recovery of modulation depth.

\begin{figure}[htb!]
\centering
\includegraphics[width=0.8\textwidth]{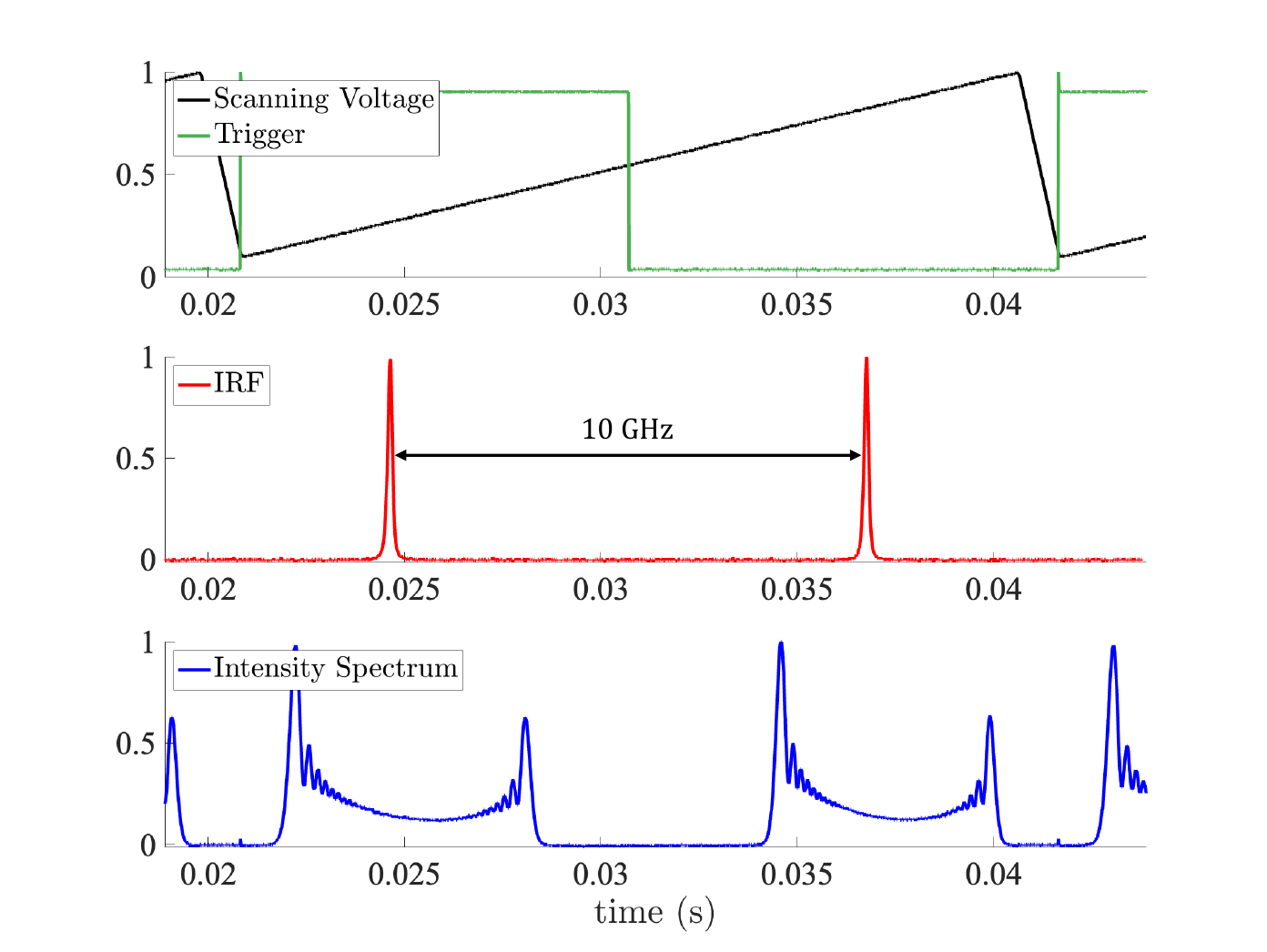}
\caption{ (Top) Linear scanning voltage sent to the piezoelectric transducer to control the cavity length and a 5V trigger. (Middle) FPI signal output when monochromatic light is directed into the FPI (no modulation). (Bottom) FPI signal output when the lasers current is modulated by a sinusoidal waveform. }
\label{fig: FPI sigs}
\end{figure}

\begin{figure}[ht!]
\centering
\includegraphics[width=0.9\textwidth]{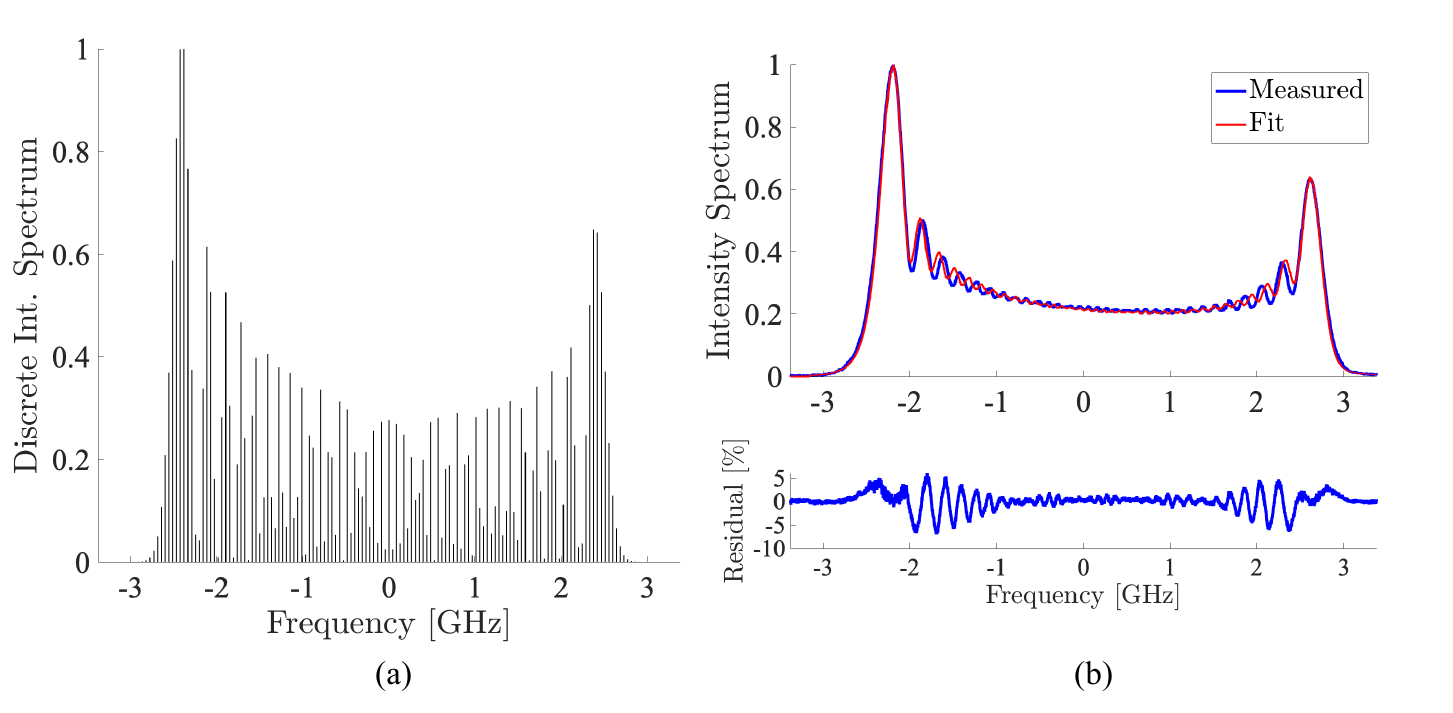}
\caption{ (a) Simulated discrete intensity spectrum for laser light with $\nu\sub{o} = 1469.3$ nm being modulated at 45 MHz with $a = 0.0845$ \un{cm^{-1}}. (b) Best fit and measured intensity spectrum with the residual shown as a percentage of the peak.}
\label{fig: FPI spec fit}
\end{figure}

Shown in Fig. \ref{fig: FPI sigs} are the signal outputs from the FPI control box (Thorlabs SA201) normalized by their peak amplitude. The FPI control box supplies the linear scanning voltage ramp to drive the piezoelectric transducer and amplifies the output from the FPI photodiode. A SA210-12B scanning FPI from Thorlabs with a NIR photodiode (1275nm - 2000nm) and a 10 GHz FSR is used. Like an etalon, a peak appears each time the FPI scans 1 FSR as shown in the middle of Fig. \ref{fig: FPI sigs}. The shape of these peaks describes how the FPI broadens monochromatic light and is referred to as the instrument response function (IRF) \cite{mathews2020near}. When the laser light is modulated at some frequency and directed into the FPI, the output signal is referred to as the intensity spectrum and is shown at the bottom of Fig. \ref{fig: FPI sigs}. Modulation depth is then recovered by simulating the discrete intensity spectrum outlined in \cite{mathews2020near}, performing a convolution with the IRF, and fitting to the normalized measured intensity spectrum. Examples of the discrete intensity spectrum and best fit to the measured intensity spectrum are shown in Fig. \ref{fig: FPI spec fit} for a TDL emitting near 1469.3 nm that was modulated at 45 MHz.

It was noted by \citeauthor{mathews2020near} \cite{mathews2020near}, that the fitting procedure to the transmitted intensity spectrum of the FPI is insensitive to the relative phase-shift $\psi_1$. To determine this parameter, the authors fit to full-spectrum scanned-WMS harmonics obtained at 1 kHz but with the same modulation frequency. For high $E''$ transitions, this requires a high-temperature static gas cell \cite{ucla_gascell} where the temperature, pressure, and gas composition are known, making $\psi_1$ the only unknown, provided all other laser tuning parameters have been determined. 

Recently, \citeauthor{guerrero2025MHExT} proposed an alternative method that is more accessible to industry and other research laboratories not equipped with such a facility \cite{guerrero2025MHExT}. It leverages the fact that the position of the absorption peaks during the up and down portions of a scan are completely determined by the relative phase shift $\psi_1$. By acquiring an absorption and baseline intensity signal at the modulation frequency of interest $\psi_1$ can be determined through a simple fitting routine. The general procedure is to first determine $\varphi$ by fitting Eq. (\ref{eq: I-fit}) to the baseline intensity signal, and then to plot absorbance versus wavenumber by generating a vector using Eq. (\ref{eq: v-fit}) where $\delta$ is a free parameter.
\begin{equation}\label{eq: I-fit}
    I_0(t) = c_1\cos(\omega t + \varphi) + c_2
\end{equation}

\begin{equation}\label{eq: v-fit}
    \nu(t) = \cos(\omega t + \delta)
\end{equation}

The two absorption peaks will superimpose onto each other once the relative phase-shift $(\varphi - \delta)$ is equal to $\psi_1$. Figure \ref{fig: psi1 char}(b) illustrates the overlapping of absorbance peaks for data obtained at 500 kHz. To asses the accuracy of this method, $\psi_1$ was determined for a TDL emitting near 1391.7 nm at several kHz rate modulation frequencies and compared to the values obtained using an etalon. Errors less than 1\% of the known values were reported \cite{guerrero2025MHExT}. The advantages of this method compared to fitting to scanned-WMS-$nf/1f$ harmonics are (1) the method is independent of other laser characterization parameters, (2) inferring $\psi_1$ does not depend on the thermodynamic properties of the environment in which the signals were acquired, and (3) the post-processing required is much simpler.


\section{Practical Considerations}




\subsection{Background Subtraction}

In general, the background and absorption signals are obtained at a random phase since data acquisition is generally triggered by an external event that can occur at any instant in time. Before gas properties can be inferred, the two signals must first be aligned to perform background subtraction as outlined in section \ref{sec: calibration-free WMS}. The procedure given here is taken from \cite{mathews2021infrared}. 

First, if the modulation frequencies are divisible by the scanning frequency as discussed in section \ref{sec: MHz rate freq}, then each scan is periodic and since the background is assumed to remain constant, then the background signal can be on the order of 10-100 scans. This can help reduce memory storage since at a sampling frequency of 3 GHz with a 12-bit resolution data acquisition card, every 1 second of data requires 4.5 GB of memory storage. Additionally, since each scan is periodic, the procedure here only needs to search one scan period to align the signals. 

\begin{figure}[h]
\centering
\includegraphics[width=0.9\textwidth]{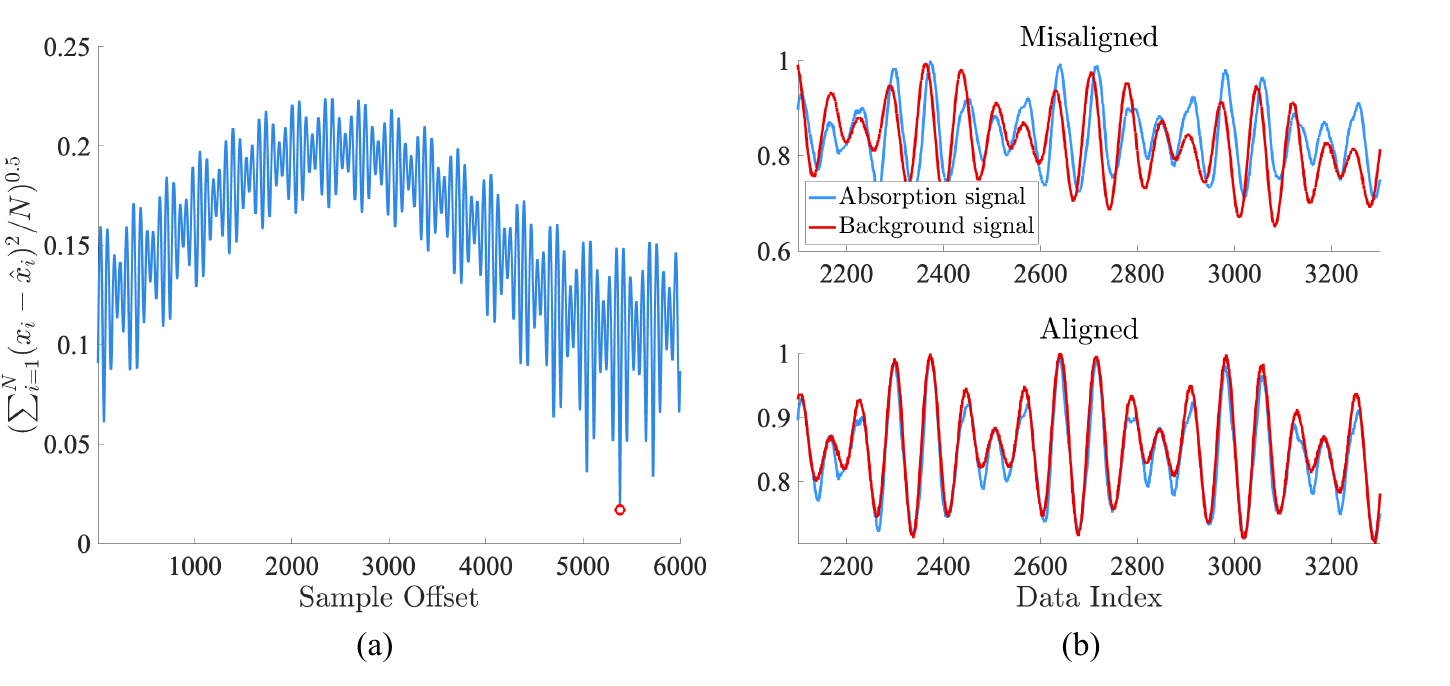}
\caption{ (a) Root-mean-square deviation as the background signal is translated. The red dot marks the minimum RMSD. (b) Example SWMS signals before and after alignment.}
\label{fig: Bg align}
\end{figure}

The procedure consists of translating the background intensity signal by 1 data point and computing the root-mean-square deviation (RMSD). This is done for a total of $f\sub{s}/f\sub{scan}$ points where $f\sub{s}$ is the sampling frequency. To improve the comparison, the two signals are normalized by their peak intensity. Shown in Fig. \ref{fig: Bg align}(a) is the RMSD of the signals at each sample offset with a red dot indicating the minimum RMSD and in Fig. \ref{fig: Bg align}(b) the misaligned and aligned signals.

\subsection{Bias-Tee Circuitry}

The demand for high-speed TDLAS sensors operating in the MHz range continues to grow, particularly for applications involving flows with short temporal scales, such as those containing detonation waves. In section \ref{sec: MHz rate freq}, the limitations on measurement rates were discussed, along with the recent adoption of bias-tee circuitry as an enabling technology for MHz-rate sensors. Although the first MHz-rate scanned-WMS sensor was demonstrated in 2020 by \citeauthor{mathews2020near} \cite{mathews2020near}, providing a foundational sensor architecture and techniques for laser characterization at MHz modulation frequencies, subsequent efforts to reproduce the sensor have faced challenges due to several differences in more recent versions of the same hardware.

At the time of writing this manuscript, all MHz-rate scanned-WMS sensors demonstrated have used laser diode mounts with integrated bias-tee circuitry (ILX Lightwave LDM-4984-BTB) \cite{mathews2020near, mathews2021high, schwartz2023near, Guerrero2025Quant, guerrero2025MHExT}. These mounts are originally designed for type-1 laser diodes, which have the anode connected to pin 10 and the cathode connected to pin 11. However, laser diodes from NTT Electronics America / NEL, Nanoplus, and Eblana photonics, commonly used suppliers, have the cathode on pin 12 and the anode on pins 11 and 13, which complicates modulation through the external bias-tee connection. Nanoplus offers a second pin-out configuration with the anode on pin 10 and and cathode on pin 9 but is still incompatible with the bias-tee LDMs. To address this, \citeauthor{guerrero2025MHExT} \cite{guerrero2025MHExT} performed the following modifications along with the custom wiring configuration summarized in Table \ref{tab: wiring}.

\begin{enumerate}
    \item The jumper labeled "B" on the printed circuit board (PCB) had to be unsoldered and the jumper labeled "C" had to be soldered instead.
    \item The pin from screw terminal 11 comes cut out of the box to prevent contact with the PCB and needs to be reconnected to the PCB.
    \item The inductor-capacitor-resistor (LCR) bridge across pins 10 and 11 needs to be placed across pins 11 and 12 in the same order.
    \item The brown wire which would normally go to the cathode, is soldered to the LD bias pad.
    \item The laser anode (pin 11 or 13) was electrically connected to the chassis ground via  a jumper cable.
\end{enumerate}

\begin{table}[h]
  \centering
  \caption{ILX Lightwave LDM-4984-BTB wiring for NTT Electronics America / NEL laser diodes.}
    \begin{tabular}{@{}lll@{}}
    \toprule
    Wire Color & Description & Connected to \\
    \midrule
    Green  & Chassis ground & - \\
    Brown  & Laser cathode & LD bias pad \\
    Blue  & PD cathode & 5 \\
    Gray  & PD anode & 4\\
    White  & Laser Anode & 11\\
    Red  & TE+ & 6\\
    Black  & TE- & 7 \\
    Orange  & Therm+ & 1 \\
    Yellow & Therm- & 2 \\
    \bottomrule
    \end{tabular}%
  \label{tab: wiring}%
\end{table}%


In addition to these PCB modifications, they found that the choice of laser diode controller (LDC) significantly impacts the stability of the laser. Through testing of three different controllers (ILX LDC-3900, ILXLDC-3726, and LDC-3908) they determined that only the LDC-3908 model with LDC-3916372 modules could effectively maintain both the DC current and temperature of the laser diode when modulation was applied through the bias-tee input. These findings suggest that the MHz-rate scanned-WMS sensor is highly dependent on specific hardware components, particularly the laser diode mounts and controller. This underscores the need for MHz-rate scanned-WMS sensor designs that are not constrained by the choice of LDC. One potentially alternative that has not been explored is the use of laser diodes with integrated bias-tee circuitry, such as those offered by Aerodiode and Eblana Photonics. This could eliminate the LDMs and allow the use of any LDC.

\section{Scanned-DA vs Scanned-WMS}

The theory and practical aspects of both scanned-DA and scanned-WMS techniques have been thoroughly presented. In this section, we briefly compare the two techniques. The most notable difference is the more complex laser characterization and data post-processing required for scanned-WMS. In environments with uniform gas properties and high SNR, post-processing of scanned-WMS data is relatively straightforward. However, in non-uniform flow fields, the transition linecenter can shift, leading to peak blending and reduced prominence of the WMS harmonic signals. In addition, side peaks may appear if the line shift is significant \cite{guerrero2025MHExT}. These effects require manual evaluation of individual peaks, which must be discarded if they (1) are blended, (2) correspond to side peaks, or (3) are only observable for one of the lasers. For MHz-rate scanned-WMS measurements, this challenge is further amplified, as in 0.1 seconds of data, there are an astounding 100,000 peaks, each of which needs to be examined, making the process significantly more time consuming. 

In contrast, scanned-DA data post-processing is significantly simpler and can be easily automated. However, at high pressures where collisional line broadening becomes prominent, identifying a non-absorbing region in the transmitted intensity signal can be challenging. This non-absorbing portion is essential for correcting the transmitted intensity to a baseline signal, as scanned-DA methods are susceptible to beam steering and transmission losses. These limitations make the WMS techniques more favorable in harsh environments, due to their superior noise rejection capabilities and the ability to correct for non-absorbing transmission losses through normalization by the $1f$ harmonic signal. Therefore, researchers must carefully assess the advantages and limitations of each method based on their specific application. A summary of these and other considerations is provided in Table \ref{tab: comparison}. 

\begin{table}[htbp]
\footnotesize
  \centering
  \caption{Comparison of basic considerations when selecting between scanned-DA and scanned-WMS techniques.}
    \begin{tabular}{p{10em}p{15em}p{18em}}
    \toprule
    \textbf{Metric} & \textbf{Scanned-DA} & \textbf{Scanned-WMS} \\
    \midrule
    Data post-processing & Straight forward, only an etalon signal is required to determine the relative wavelength tuning curve. & More complex, requiring careful characterization of the laser tuning parameters, and simulation of look up tables. \\ [2ex]
    Sampling rate & Sampling frequencies of 250 - 500 MHz have been used to acquire scanned-DA data at 1 MHz. \cite{nair2020mhz}  & Sampling frequencies of 1 - 3 GHz have been used to acquire scanned-WMS data at 1 MHz. \cite{mathews2021infrared,Guerrero2025Quant}  \\[1ex]
    Wavelength characterization & An etalon may be used at both kHz and MHz rates. & At MHz rates, a FPI is required to determine modulation depth. \cite{mathews2020near} \\[1ex]
    Detection limits & Generally, this method works well at peak absorbance levels between 0.1 - 1 \cite{nair2020mhz,kuenning2024multiplexed,minesi2022multi}& The noise rejection capabilities of this method make it ideal for absorbance environments where the peak absorbance is less than 0.1 \cite{rieker2009calibration}. High absorption levels can also cause distortion in the WMS harmonics. \cite{goldenstein2014fitting}. \\[1ex]
    Challenges in harsh environments & A non-absorbing portion in the transmitted intensity signal is required to correct the signal for non-absorbing losses. \cite{spearrin2014simultaneous} At high-pressures sufficient scan-depth is required to include the wings of the absorption feature \cite{nair2022extended,nair2023optical}. & Due to normalizing by the $1f$ harmonic signal, the WMS-$2f/1f$ harmonic become insensitive to non-absorbing losses. Broadening of lines is also not a challenge since a baseline signal is not required. \cite{rieker2009calibration, goldenstein2014fitting} \\[10ex]
    Bias-tee circuity & Relatively simple to implement with no hardware specific requirements & Currently demonstrated MHz-rate scanned-WMS sensors rely on specific hardware and laser diode pin-out configurations. \cite{guerrero2025MHExT} \\[2ex]
    Time requirements & Low. Only the characterization of the lasers wavelength tuning  range is required. Post-processing can also be automated \cite{kuenning2024multiplexed, nair2020mhz}. & For kHz rate WMS, medium. For MHz rate WMS high.  Careful characterization of the laser's tuning parameters takes significantly more time, especially at MHz rates, compared to scanned-DA. Additionally, in harsh environments with short path lengths, peak labeling can be extremely time consuming, particularly at MHz rates, due to the large number of peaks that require manual identification. \\[1ex]
    Initial cost  & Medium, primarily the cost of the laser diodes, etalon, and detectors. & High due to the additional purchasing of a DAQ capable of the high sampling rates required and due to the new instrument required (FPI) to characterize modulation depth. \\
    \bottomrule
    \end{tabular}%
  \label{tab: comparison}%
\end{table}%

\section{Conclusions}

The theory of gas absorption was mathematically presented, incorporating key concepts from classical optics and, where necessary, quantum mechanics to enhance understanding. At this point, the reader should have a clear understanding of units related to absorption spectroscopy, and of the different conventions for the Beer-Lambert Law that are commonly encountered. They should also have a clear understanding of broadening mechanisms and the resulting lineshapes. Additionally, best practices for obtaining quantitative measurements using the scanned-direct absorption method, challenges associated with optical pressure diagnostics, general rules for line selection, and the interpretation of gas properties derived from TDLAS in non-uniform environments should be well understood.

Related to scanned-WMS, the calibration-free WMS model was derived with more detail than found in other publications and the shape of the WMS harmonics was related to the derivatives of the spectral lineshape. Methods for making quantitative measurements were also discussed with a focus on the two-color ratio of WMS-$2f/1f$ signals and its use in thermometry, and on the WMS-$4f/2f$ signal and how it can be used to measure collisional width in a bath gas of unknown composition. Furthermore, the need for MHz rate modulation was demonstrated by examining the maximum allowed sensor measurement rate for a pair of modulation frequencies. Lastly, practical aspects of implementing a SWMS sensor were discussed including laser characterization at kHz and MHz modulation frequencies, background subtraction, and debugging of critical hardware used in MHz rate SWMS sensors.





\bibliographystyle{elsarticle-num-names}

\bibliography{TDLcitations,RDEcitations}







\end{document}